\documentclass[]{article}
\usepackage{graphicx, amsmath, tikz, graphicx, cite, xcolor, braket, caption, subcaption} 
\usepackage[margin = 2cm]{geometry}
\usepackage[colorlinks = true]{hyperref}
\usepackage{ulem}
\title{QUANTUM STATE TRANSFER AND PERIODICITY IN DISCRETE-TIME QUANTUM WALKS UNDER NON-MARKOVIAN DEPHASING NOISE}
\author{
	Monika Rani$^1$\thanks{Email: \texttt{rani.2@iitj.ac.in}}, Supriyo Dutta$^2$\thanks{Email: \texttt{dosupriyo@gmail.com}}, Subhashish Banerjee$^1$\thanks{Email: \texttt{subhashish@iitj.ac.in}}\\
	\small{$^1$Department of Physics, Indian Institute of Technology Jodhpur,} \\
	\small{Jodhpur, Rajasthan, India-342037.}\\
	\small{$^2$Department of Mathematics, National Institute of Technology Agartala,} \\
	\small{Jirania, West Tripura, India - 799046.}
}
\date{} 

\begin{document}
	
	\maketitle
	
	\begin{abstract}
		 In quantum communication, quantum state transfer from one location to another in a quantum network plays a prominent role, where the impact of noise could be crucial. The idea of state transfer can be fruitfully associated with quantum walk on graphs. We investigate the consequences of non-Markovian quantum noises on periodicity and state transfer induced by a discrete-time quantum walk on graphs, governed by the Grover coin operator. Different bipartite graphs, such as the path graph, cycle graph, star graph, and complete bipartite graph, present periodicity and state transfer in a discrete-time quantum walk depending on the topology of the graph. We investigate the effect of quantum non-Markovian dephasing noises, particularly quantum non-Markovian Random Telegraph Noise (RTN) and modified non-Markovian Ornstein-Uhlenbeck Noise (OUN) on state transfer and periodicity. We demonstrate how the RTN and OUN noises allow state transfer and periodicity for a finite number of steps in a quantum walk. Our investigation brings out the feasibility of state transfer in a noisy environment.\\

	\end{abstract}
	
	\tableofcontents
 
	\newpage
	\section{Introduction}
		
		There are several desired characteristics of the large-scale quantum computation \cite{wilde2013quantum, bennett1998quantum}. The transmission and manipulation of quantum states without destroying their fragile coherence is of primary importance among them. Originally, the transport of quantum information was proposed by using quantum entanglement to communicate the information between atomic clouds over long distances, which results in quantum teleportation of states. Quantum state transfer \cite{bose2003quantum, christandl2004perfect} between different spins is another aspect of quantum information transfer. Faithful transfer of quantum information between different locations of the same quantum hardware is a fundamental building block of several quantum information processing protocols. In all quantum information-theoretic tasks, quantum noise is inevitable \cite{banerjee2018open}. Therefore, quantum state transfer between different locations of a quantum network under noise is an important topic for investigation.
	
	  	In quantum information and computation, quantum walks \cite{aharonov1993quantum, konno2008quantum, venegas2012quantum, kempe2003quantum} serve as a universal framework for quantum computation \cite{childs2009universal} and have been employed to devise numerous quantum algorithms \cite{xia2019random, dutta2023quantum}, such as, quantum search algorithms, element distinctness \cite{ambainis2007quantum}, Boolean formula evaluation \cite{farhi2007quantum}, ranking the vertices of a network \cite{paparo2012google, dutta2025discrete, chawla2020discrete} and many others. Broadly, quantum walks can be classified into two categories: discrete-time and continuous-time quantum walks \cite{chandrashekar2010relationship, chandrashekar2007symmetries, childs2010relationship, venegas2012quantum}. Here, we use the discrete-time quantum walk (DTQW). DTQW is a valuable tool in quantum computing and information theory, offering a rich framework for exploring quantum systems. Quantum walk and its relation with state transfer are well-studied in the literature \cite{kurzynski2011discrete, vstefavnak2016perfect, santos2022quantum, kendon2011perfect, dutta2022perfect}. Let a quantum walker begin walking from the vertex $u$ of a network. If, after time $\tau$, the walker is at vertex $v$ with probability $1$, we say that the state of the vertex $u$ is transferred to the vertex $v$ at time $\tau$. If the initial and the final vertices are the same, that is, $u = v$, the walk is periodic at vertex $u$ at time $\tau$. This phenomenon is coined as the periodicity of the quantum walk. 
        
	  To analyze state transfer and periodicity in quantum walks, fidelity is used to compare the evolved state with the receiver state over time. Fidelity is a measure of similarity between two quantum states, ranging from 0 (indistinguishable) to 1 (distinguishable). It helps quantify how accurately a quantum state has evolved or transferred to a desired state. The importance of quantum state transfer lies in its role in preserving quantum coherence and entanglement during quantum teleportation \cite{sherson2006quantum, ren2017ground, bouwmeester1997experimental}, entanglement distribution \cite{cirac1997quantum} and quantum error correction \cite{kay2016quantum}. Efficient and accurate state transfer is crucial for the development of quantum communication protocols and distributed quantum networks. The perfect state transfer is a rare incident when the quantum network is represented by a simple graph.
      
      DTQWs provide a powerful framework for implementing quantum state transfer on graphs. A crucial element in these walks is the choice of the coin operator, which governs the walker’s movement across different states. Different types of coin operators have been studied in the literature, such as the Grover coin operator \cite{mukai2020discrete}, the Fourier Coin operator \cite{jayakody2021one, mukai2020discrete} and the Hadamard coin operator \cite{chandrashekar2007symmetries, jayakody2021one}. The Grover operator, originally devised for quantum search algorithms \cite{grover1996fast}, presents unique characteristics that make it a compelling choice for the coin operator, especially in higher-dimensional spaces and is made use of in this work. State transfer based on a DTQW has been studied on different classes of bipartite graphs, such as complete bipartite graphs \cite{vstefavnak2017perfect}, star graphs \cite{santos2022quantum}, trees \cite{dimcovic2011framework}, and a few other graphs \cite{barr2012periodicity, zhan2019infinite, yalccinkaya2015qubit, kurzynski2011discrete}. The structure of the graph significantly influences the dynamics of state transfer. 
	    
		Noise is inevitable in nature. In realistic quantum systems, environmental noise introduces decoherence, which can degrade coherence and fidelity during the state transfer. Interestingly, certain types of non-Markovian noise characterized by memory effects can counteract this degradation by restoring lost coherence. In particular, the environmental noise can be a significant issue in the scaling up of the number of steps in a quantum walk system. The effect of quantum noise on continuous-time quantum walk has been investigated in \cite{benedetti2016non, rossi2017continuous}. In \cite{rossi2017continuous}, the effect of noise on localization properties of quantum states was investigated on lattice graphs. The discrete-time quantum walks on different graph topologies, such as cycle graphs, path graphs, under a number of quantum noises were investigated in  \cite{chandrashekar2010relationship, chandrashekar2007symmetries, banerjee2017non}. A new model of discrete-time quantum walk in a noisy environment was presented in \cite{rani2024non}, and is applicable for any graph topology. Understanding the characteristics of the noisy and noiseless quantum walk could assist in the further exploration of possible systems, where the quantum walk is an essential component. Quantum state transfer and periodicity are an important feature of quantum walk. Therefore, it would be pertinent to consider state transfer and periodicity on a DTQW in a noisy environment. There are different types of noise considered in the literature, such as dephasing noise \cite{shrikant2018non, rani2024non, dutta2023qudit}, such as random telegraph noise, and Ornstein-Uhlenbeck noise \cite{kumar2018non, banerjee2017non}, as well as the amplitude damping noise \cite{srikanth2008squeezed, rani2024non, dutta2023qudit}, in both the  Markovian and non-Markovian scenarios. Here, we use the non-Markovian RTN and modified non-Markovian OUN noises generalized to arbitrary dimensions on a quantum walk. These noises were originally defined to qubits but can be generalized to higher-dimensional systems using Weyl operators \cite{dutta2023qudit}. There are various proposals to apply quantum noise to quantum walks. In our model, we apply quantum noise to the state of the walker after a finite number of steps of the quantum walk \cite{kumar2018non}. 

         The state transfer depends on the topology of the graphs. In a path graph, the walker can only move step by step along the line of vertices, as the topology of the path graph is linear. In a cycle, the topology is circular \cite{banerjee2008symmetry}. In a star graph, the central vertex plays a crucial role, making the state transfer highly dependent on whether the sender or receiver is at the central vertex or the external vertices. In a complete bipartite graph, the topology consists of two sets of vertices, where every vertex in one set is connected to all the vertices in another set. In this case, we observe that maximum fidelity in state transfer occurs when both the sender and receiver are within the same sets. In contrast, when the sender and receiver are placed in opposite sets, the fidelity is reduced. We observe a number of noise effects on the fidelity:
		\begin{enumerate}
			\item 
				For the path graph, state transfer and periodicity do not depend on the noise but depend on the positions of the sender and receiver of quantum information. In the case of state transfer under noise,  maximum fidelity occurs when the sender and receiver are located at the end position of the path graph.
			\item 
				Noise reduces fidelity very little when the sender and receiver are on the opposite side of the cycle graph due to constructive interference. There is no effect of noise when the sender and receiver are at asymmetrical places in the cycle graph, but there is a sharp drop in fidelity.
			\item 
				There is no effect of noise on the state transfer on the star graph when the sender is at the central vertex and the receiver is at any external vertex of the graph. But the fidelity of state transfer starts to decrease when the sender and receiver positions interchange. Periodicity shows an oscillating behaviour under non-Markovian RTN noise, while a damping nature under non-Markovian OUN noise with time steps is observed.
			\item 
				The fidelity of state transfer and periodicity under noise demonstrates an oscillatory nature with time steps on the complete bipartite graph.
		\end{enumerate}
	
	This article is distributed as follows. In Section 2, we present the preliminary ideas on graph theory and quantum walks. Section 3 describes the general construction of the shift and coin operators to define the DTQW on a graph. We discuss the non-Markovian channels in the next section. Section 5 consists of different subsections which describe the state transfer and periodicity on the considered graphs. Finally, we make our conclusion. 
	
	\section{Preliminary}  
	
	A graph $G = (V(G), E(G))$ is a combinatorial object 
	made up of a set of vertices $V(G)$ and a set of edges $E(G)$, where each edge connects two vertices \cite{west2001introduction, bondy1976graph, dutta2016graph}. The number of vertices in a graph is the order of the graph, which is denoted by $n$ throughout this article. An edge $(u, v)$ connects two vertices $u$ and $v$, which represents an interaction between these vertices. We say that $u$ and $v$ are adjacent to each other. Also, the edge $(u, v)$ is incident to the vertices $u$ and $v$. A directed edge has a direction on it. A directed edge from the vertex $u$ to the vertex $v$ is denoted by $\overrightarrow{(u, v)}$. The directed edge $\overrightarrow{(u, v)}$ is said to be outgoing from $u$ and incoming to $v$. A graph is said to be undirected if it has no directed edges. All the edges in a directed graph are directed. A loop is an edge joining a vertex with itself. A simple graph is an undirected graph having no loops. Also, in a simple graph, there is at most one edge joining two vertices. We do not consider any weight on the edges \cite{adhikari2017laplacian}. We can convert a simple graph $G$ to a directed graph $\overrightarrow{G}$ by assigning two opposite directions on each undirected edge. Therefore, corresponding to an undirected edge $(u, v)$ in $G$, there are two directed edges $\overrightarrow{(u, v)}$ and $\overrightarrow{(v, u)}$ in $\overrightarrow{G}$.
	
	The degree of a vertex $v$ in an undirected graph $G$ is the number of edges incident on it, which is denoted by $d_v$. In a directed graph $\overrightarrow{G}$, the number of directed edges coming into a vertex $v$ is called the indegree of $v$, denoted by $d_v^{(i)}$. Similarly, the number of edges going out from a vertex $u$ is called the outdegree of $u$, denoted by $d_u^{(o)}$. The degree of an undirected graph $G$ is $d(G) = \sum_{v \in V(G)} d_v$ throughout this article. 
	
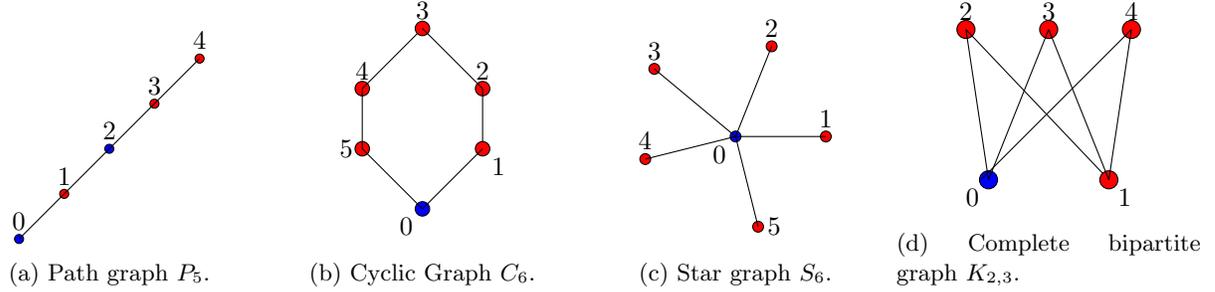
\begin{figure}
		\centering
		\begin{subfigure}{0.23\textwidth} 
			\centering
			\begin{tikzpicture}[scale=.6] 
			\draw [fill = blue] (-2, -2) circle [radius = 1mm];
			\node [above] at (-2, -2) {$0$};
			\draw [fill = red] (-1,-1) circle [radius = 1mm];
			\node [above] at (-1, -1) {$1$};
			\draw [fill = blue] (0, 0) circle [radius = 1mm];
			\node [above] at (0, 0) {$2$};
			\draw [fill = red] (1, 1) circle [radius = 1mm];
			\node [above] at (1, 1) {$3$};
			\draw [fill = red] (2, 2) circle [radius = 1mm];
			\node [above] at (2, 2) {$4$};
			\draw (-2, -2) -- (-1, -1);
			\draw (-1, -1) -- (0, 0);
			\draw (0,0) -- (1, 1);
			\draw (1, 1) -- (2, 2);
			\end{tikzpicture}
			\caption{Path graph $P_{5}$.}
			\label{fig:path_graph}
		\end{subfigure}
		\begin{subfigure}{0.23\textwidth}
			\centering
			\begin{tikzpicture}[scale=.8] 
			\draw [fill = blue] (0, 0) circle [radius = 1.2mm];
			\node [below left] at (0, 0) {$0$};
			\draw [fill = red] (1, 2) circle [radius = 1.2mm];
			\node [above] at (1, 2) {$2$};
			\draw (1, 1) -- (1, 2);
			\draw [fill = red] (1,1) circle [radius = 1.2mm];
			\node [below right] at (1, 1) {$1$};
			\draw (0, 0) -- (1, 1);
			\draw [fill = red] (-1, 2) circle [radius = 1.2mm];
			\node [above] at (-1, 2) {$4$};
			\draw (0, 3) -- (-1, 2);
			\draw [fill = red] (-1, 1) circle [radius = 1.2mm];
			\node [left] at (-1, 1) {$5$};
			\draw (-1, 2) -- (-1, 1);
			\draw (0, 0) -- (-1, 1);
			\draw [fill = red] (0, 3) circle [radius = 1.2mm];
			\node [above] at (0, 3) {$3$};
			\draw (1, 2) -- (0, 3);
			\end{tikzpicture}
			\caption{Cyclic Graph $C_{6}$.}
			\label{fig:cycle_graph}
		\end{subfigure}
		\begin{subfigure}{0.23\textwidth}
			\centering
			\begin{tikzpicture}[scale=.6] 
			\draw [fill = blue] (0, 0) circle [radius = 1.2mm];
			\node [below left] at (0, 0) {$0$};
			\draw [fill = red] (0.8, 2) circle [radius = 1.2mm];
			\node [above] at (0.8, 2) {$2$};
			\draw (0, 0) -- (0.8, 2);
			\draw [fill = red] (2,0) circle [radius = 1.2mm];
			\node [above] at (2, 0) {$1$};
			\draw (0, 0) -- (2, 0);
			\draw [fill = red] (-2, -0.5) circle [radius = 1.2mm];
			\node [above] at (-2, -0.5) {$4$};
			\draw (0, 0) -- (-2, -0.5);
			\draw [fill = red] (.5, -2) circle [radius = 1.2mm];
			\node [right] at (.5, -2) {$5$};
			\draw (0, 0) -- (.5, -2);
			\draw [fill = red] (-1.8, 1.5) circle [radius = 1.2mm];
			\node [above] at (-1.8, 1.5) {$3$};
			\draw (0, 0) -- (-1.8, 1.5);
			\end{tikzpicture}
			\caption{Star graph $S_{6}$.}
			\label{fig:star_graph}
		\end{subfigure}
		\begin{subfigure}{0.23\textwidth}
			\centering
			\begin{tikzpicture}[scale=1]
			\draw [fill = blue] (-.4, 0) circle [radius = 1.2mm];
			\node [below left] at (-.4, 0) {$0$};
			\draw [fill = red] (-0.7, 2) circle [radius = 1.2mm];
			\node [above] at (-0.7, 2) {$2$};
			\draw (-.4, 0) -- (-0.7, 2);
			\draw (1.2, 0) -- (-0.7, 2);
			\draw [fill = red] (1.2,0) circle [radius = 1.2mm];
			\node [below right] at (1.2, 0) {$1$};
			\draw [fill = red] (1.5, 2) circle [radius = 1.2mm];
			\node [above] at (1.5, 2) {$4$};
			\draw (-.5, 0) -- (1.5, 2);
			\draw (1.2, 0) -- (1.5, 2);
			\draw [fill = red] (0.4, 2) circle [radius = 1.2mm];
			\node [above] at (0.4, 2) {$3$};
			\draw (-.4, 0) -- (0.4, 2);
			\draw (1.2, 0) -- (0.4, 2);
			\end{tikzpicture}
			\caption{Complete bipartite graph $K_{2,3}.$}
			\label{fig:complete_bi_graph}
		\end{subfigure}
		\caption{Examples of simple graphs, which are used in this article.}
		\label{simple_graphs}
	\end{figure}

	In this work, we use several simple graphs, such as path graphs, cycle graphs, star graphs, complete bipartite graphs, etc., which are depicted in figures \ref{fig:path_graph}, \ref{fig:cycle_graph}, \ref{fig:star_graph}, and \ref{fig:complete_bi_graph}, respectively. Let a graph have $n$ vertices $ 0, 1, \dots (n - 1)$. A path graph $P_n$ has edges $\{(0, 1), (1, 2), \dots ((n-2), (n-1))\}$. Therefore, in $P_n$ the degree of vertices $0$ and $(n-1)$ is $1$ and all other vertices have degree $2$. A cycle graph $C_n$ has edges $\{(0, 1), (1, 2), \dots ((n-1), 0)\}$. Therefore, every vertex in $C_n$ has degree $2$. In a star graph $S_n$, the external vertices $1, 2, \dots {(n - 1)}$ are adjacent to the central vertex $0$, only. Therefore, the degree of $0$ is $(n-1)$, and the degree of all other vertices is $1$. The vertex set of a complete bipartite graph $K_{m, n}$ has two partitions $V_1$ and $V_2$ having $m$ and $n$ vertices, respectively. Also, it contains all possible edges with one end in $V_1$ and another end in $V_2$. 
	
	In quantum information theory, a pure quantum state is a normalized vector which is denoted by $\ket{\bullet}$. A quantum state can also be represented by a density matrix, which is a positive semidefinite Hermitian matrix with unit trace. Corresponding to the pure quantum state $\ket{\psi}$, there is a density matrix $\rho = \ket{\psi}\bra{\psi}$. We utilize the Kraus operators \cite{nielsen2010quantum, kraus2005operations} to represent a quantum channel. A set of Kraus operators $\{K_{i}\}$ satisfies the completeness relation which is 
	\begin{equation}\label{Kraus_operator_general}
	\sum_i K_i^\dagger K_i = I, 
	\end{equation}
	where $I$ is the identity operator.	Let $\rho$ be the density matrix of a quantum state which we transfer via a channel represented by the set of Kraus operators $\{K_{i}\}$. Then the final state is given by  
	\begin{equation}
	\rho' = \sum_{i} K_{i} \rho K_{i}^\dagger.
	\end{equation}
	We utilize the Weyl operators \cite{bertlmann2008bloch, weyl1927quantenmechanik} to construct the Kraus operators for arbitrary dimension \cite{dutta2023qudit, basile2024weyl}. The Weyl operators of order $d$ are represented by 
	\begin{equation}\label{Weyl_operator}
	U_{u,v}= \sum_{k=0}^{d-1} e^{(\frac{2\pi i}{d})ku} \ket{k}\bra{(k+v) \bmod d}, ~\text{where}~ 0 \leq u, v \leq (d-1).
	\end{equation}
	We can prove that $U_{u,v}$ is unitary for all $v$ and $u$, that is $U_{u,v}^\dagger U_{u,v} =  U_{u,v} U_{u,v}^\dagger = I_d$. When $u = v = 0$, we have $U_{0, 0} = I_d$. If $d = 2$ for $u = 1$ and $v = 0$, we have $U_{1,0} = \sigma_z$, which is the Pauli $Z$ matrix.
	
	To investigate quantum state-transfer, we examine revivals in the fidelity between a pair of quantum states. The fidelity between two given state vectors $\ket{\psi}$ and $\ket{\phi}$ is defined by
	\begin{equation}\label{fidelity_vectors}
	F(\ket{\psi}, \ket{\phi}) = \left( |\braket{\psi | \phi}|\right)^2.
	\end{equation}
	When two quantum states are represented by density matrices $\rho$ and $\sigma$, fidelity is determined by
	\begin{equation}\label{fidelity_density}
	F(\rho, \sigma) = \left( \text{Tr} \left[ \sqrt{ \sqrt{\rho} \sigma \sqrt{\rho} } \right] \right)^2.
	\end{equation}
	In any case, $ 0 \leq F \leq 1 $. 
	
	\section{DTQW on graphs}
	
      DTQW \cite{banerjee2008symmetry, chandrashekar2007symmetries} is defined on a graph where a walker can move from one vertex to the other vertices following the laws of quantum mechanics. In the case of the DTQW, the state of the walker is described by a vector in a Hilbert space and the evolution of the system is governed by a unitary operator. In this work, we consider the discrete-time coined quantum walk. The unitary operator is a product of a shift operator and a coin operator.

      Recall that a regular graph is a graph having equal degree for all its vertices. In our work, the cycle graphs is the only regular graphs. All the other graphs are irregular. We need a framework which is applicable for all. Therefore, we convert all the undirected graphs to directed graphs. Then we define a quantum walk on it. This method is more efficient than the conventional vertex-based approach of quantum walk, which is limited to regular graphs.
	
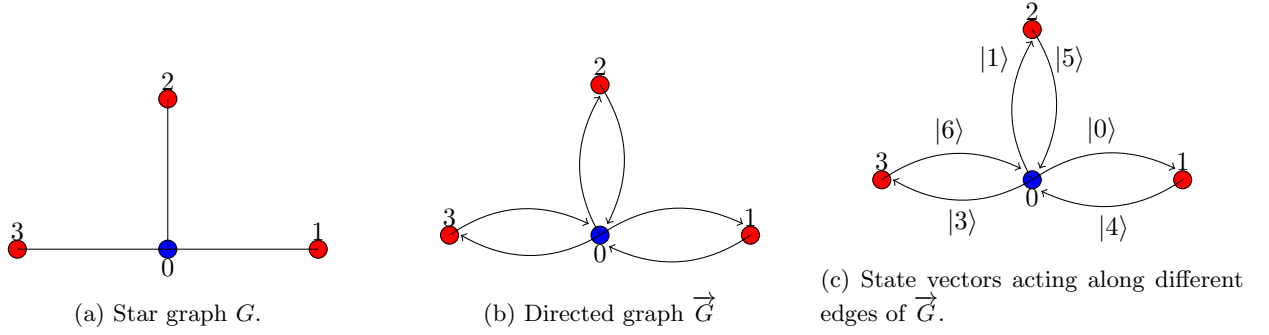
\begin{figure}
		\centering
		\begin{subfigure}{0.32\textwidth} 
			\centering
			\begin{tikzpicture}[scale=1] 
			\draw [fill = blue] (0, 0) circle [radius = 1.2mm];
			\node [below] at (0, 0) {$0$};
			\draw [fill = red] (0, 2) circle [radius = 1.2mm];
			\node [above] at (0, 2) {$2$};
			\draw (0, 0) -- (0, 2);
			\draw [fill = red] (2,0) circle [radius = 1.2mm];
			\node [above] at (2, 0) {$1$};
			\draw (0, 0) -- (2, 0);
			\draw [fill = red] (-2, 0) circle [radius = 1.2mm];
			\node [above] at (-2, 0) {$3$};
			\draw (0, 0) -- (-2, 0);
			\end{tikzpicture}
			\caption{Star graph $G$.}
			\label{star_4}
		\end{subfigure}
		\begin{subfigure}{0.32\textwidth} 
			\centering
			\begin{tikzpicture}[scale=1]
			\coordinate (v0) at (0, 0);
			\coordinate (v2) at (0, 2);
			\coordinate (v1) at (2, 0);
			\coordinate (v3) at (-2, 0);
			
			\draw[fill = blue] (v0) circle [radius = 1.2mm];
			\node [below] at (v0) {$0$};
			
			\draw[fill = red] (v2) circle [radius = 1.2mm];
			\node [above] at (v2) {$2$};
			
			\draw[fill = red] (v1) circle [radius = 1.2mm];
			\node [above] at (v1) {$1$};
			\draw[fill = red] (v3) circle [radius = 1.2mm];
			\node [above] at (v3) {$3$};
			\path[->, bend left] (v0) edge node[midway,above] {} (1.9, .15);
			\path[->, bend left] (v1) edge node[midway,below] {} (.12, -.15);
			\path[->,  bend left] (v0) edge node[midway,right] {} (0, 1.85);
			\path[->,  bend left] (v2) edge node[midway,left] {} (0.1, 0.15);
			\path[->,  bend left] (v0) edge node[midway,above] {} (-1.85, 0.001);
			\path[->,  bend left] (v3) edge node[midway,below] {} (-.15, 0.15);
			\end{tikzpicture}
			\caption{Directed graph $\overrightarrow{G}$}
			\label{directed star_4}
		\end{subfigure}
		\begin{subfigure}{0.32\textwidth} 
			\centering
			\begin{tikzpicture}[scale=1]
			\coordinate (v0) at (0, 0);
			\coordinate (v2) at (0, 2);
			\coordinate (v1) at (2, 0);
			\coordinate (v3) at (-2, 0);
			
			\draw[fill = blue] (v0) circle [radius = 1.2mm];
			\node [below] at (v0) {$0$};
			
			\draw[fill = red] (v2) circle [radius = 1.2mm];
			\node [above] at (v2) {$2$};
			
			\draw[fill = red] (v1) circle [radius = 1.2mm];
			\node [above] at (v1) {$1$};
			\draw[fill = red] (v3) circle [radius = 1.2mm];
			\node [above] at (v3) {$3$};
			\path[->, bend left] (v0) edge node[midway,above] {$\ket{0}$} (1.9, .15);
			\path[->, bend left] (v1) edge node[midway,below] {$\ket{4}$} (.12, -.15);
			\path[->,  bend left] (v0) edge node[midway,left] {} (0, 1.85);
			\path[->,  bend left] (v2) edge node[midway, right] {} (0.1, 0.15);
			\path[->,  bend left] (v0) edge node[midway, below] {$\ket{3}$} (-1.85, 0.001);
			\path[->,  bend left] (v3) edge node[midway, above] {$\ket{6}$} (-.15, 0.15);
			\node at (-.5, 1.6) {$\ket{1}$};
			\node at (.5, 1.6) {$\ket{5}$};
			\end{tikzpicture}
			\caption{State vectors acting along different edges of $\overrightarrow{G}$.}
			\label{directed_star_with_edge_states}
		\end{subfigure}
		\caption{ We consider a star graph $G$ with four vertices and three undirected edges, in sub-figure \ref{star_4}. In sub-figure \ref{directed star_4}, we assign two opposite orientations on every edge for converting it to a directed graph $\overrightarrow{G}$
		 having six directed edges. Corresponding to the directed edges, we assign a state vector from the computational basis of $\mathcal{H}^{(6)}$. The state vectors corresponding to the directed edges are marked in sub-figure \ref{directed_star_with_edge_states}.
		}
		\label{example_graph}	
	\end{figure}

	Let $G$ be a simple graph with $n$ vertices and $m$ edges. We can convert it to a directed graph $\overrightarrow{G}$ by assigning two opposite orientations on every edge. Therefore, the directed graph $\overrightarrow{G}$ has $2m$ edges. Let the vertices be labelled by $0, 1, 2, \dots (n-1)$. It assists us in arranging the directed edges of $\overrightarrow{G}$ in a lexicographic order that is
	\begin{equation}\label{lexicographic_order} 
	\overrightarrow{(u, v)} < \overrightarrow{(p, q)} ~\text{if}~ \begin{cases} & u < p, ~\text{or} \\ & v < q ~\text{and}~ u = p. \end{cases} 
	\end{equation} 
	Corresponding to every directed edge $\overrightarrow{(u, v)}$ we assign a state vector $\ket{\overrightarrow{(u, v)}}$ of the computational basis of $\mathcal{H}^{2m}$. The $k$-th edge of the lexicographic order will be assigned to the state vector $\ket{k}= (0, 0, \dots 0, 1(k\text{-th position}), 0, \dots, 0)^\dagger$ of dimension $2m$ where $k = 0, 1, \dots (2m - 1)$. The relation between directed edges and the corresponding state vector is discussed in Figure \ref{example_graph} for a star graph. 
	
	In this article, we discuss state transfer and periodicity in a DTQW. Let a sender and a receiver be placed on the vertices $s$ and $r$ of the graph $\overrightarrow{G}$, respectively. Let degree of $s$ and $r$ be $d_s$ and $d_r$, respectively, in the graph $G$. The outgoing edges from vertex $s$ can be written as $\overrightarrow{(s, v_0)}, \overrightarrow{(s, v_1)}, \dots \overrightarrow{(s, v_{d_s - 1})}$. We consider the initial state of the walker as
	\begin{equation}\label{sender_state}
	\ket{\psi_s} = \frac{1}{\sqrt{d_s}}\sum_{i = 0}^{d_s - 1} \ket{\overrightarrow{(s, v_i)}}
	\end{equation} 
	Similarly, let the incoming edges to the vertex $r$ be $\overrightarrow{(u_0, r)}, \overrightarrow{(u_1, r)}, \dots \overrightarrow{(u_{d_r-1}, r)}$, in the directed graph $\overrightarrow{G}$. If the walker reaches the vertex $r$, its state should be
	\begin{equation}\label{receiver_state}
	\ket{\psi_r} = \frac{1}{\sqrt{d_r}}\sum_{i = 0}^{d_r - 1} \ket{\overrightarrow{(u_i, r)}}.
	\end{equation}
	Clearly, $\ket{\psi_s}$ and $\ket{\psi_r}$ belongs to $\mathcal{H}^{2m}$.
	
	The shift operator in a DTQW governs the movement of the walker on the edges of the graph. The shift operator $S$, which acts on every edge of the graph, is defined by
	\begin{equation}\label{shift}
	S\ket{\overrightarrow{(i,j)}} = \ket{\overrightarrow{(j,i)}}.
	\end{equation}
	The operator $S$ swaps the direction of the walker from the edge $\overrightarrow{(i,j)}$ to $\overrightarrow{(j, i)}$. Note that, order of $S$ is $2m$ as the dimension of the Hilbert space corresponding to the edge set is $2m$.
	
	Also, corresponding to every vertex $i$, we consider a Hilbert space of $\mathcal{H}^{d_i}$ of dimension $d_i$, where $d_i$ is the degree of vertex $i$ in the undirected graph $G$. The space $\mathcal{H}^{d_i}$ is spanned by the vectors $\ket{0}, \ket{1}, \dots \ket{i}, \dots \ket{d_i - 1}$. Corresponding to every vertex $i$ with degree $d_i$ in the undirected graph $G$, we construct a Grover diffusion operator which is
	\begin{equation}\label{Grover}
	\begin{split}
	& G_i = 2 \ket{\phi(i)}\bra{\phi(i)} - I_{d_i}, ~\text{where}~ \ket{\phi(i)} = \frac{1}{\sqrt{d_i}} \sum_{j = 0}^{d_i - 1} \ket{j}.
	\end{split}
	\end{equation}
	Here, $\ket{j} = (0, 0, \dots 0, 1 (j$-th position)$, 1, \dots 0)$ is $d_i$ dimensional vector with $j = 0, 1, \dots (d_i -1)$. Now, we define the coin operator $C$ as 
	\begin{equation}\label{coin}
	C = G_0 \bigoplus G_2 \bigoplus \dots \bigoplus (- G_r) \bigoplus \dots \bigoplus (- G_s) \bigoplus \dots \bigoplus G_{n - 1},
	\end{equation}
	where $\bigoplus$ denotes the direct sum of matrices. We multiply $-1$ with the Grover operators corresponding to the sender and the receiver vertices in the expression of $C$ \cite{santos2022quantum}. This model has a number of differences with the other constructions of the coin operator \cite{vstefavnak2016perfect, vstefavnak2017perfect}. Note that the coin operator $C$ is the square matrix of order $2m$, which is equal to the sum of all vertex degrees. As the vertices in a graph may have different degrees, the Hilbert spaces corresponding to the vertices have different dimensions. Here, $C$ is defined as a direct sum, such that $G_{i}$ acts locally on the space corresponding to the vertex $i$. 
	
	Now, we are in the position to define the operator leading the DTQW on graphs, which is the product of the coin and shift operators
	\begin{equation}
	U = S \times C.
	\end{equation} 
	As the walker starts walking from the vertex $s$, the initial state of the walker is $\ket{\psi_s}$, which is defined in equation \eqref{sender_state}. The initial density matrix can be expressed as $\rho_s = \ket{\psi_s}\bra{\psi_s}$. The state of the quantum walker after $t$ steps of the quantum walk is represented by		
	\begin{equation}
	\ket{\psi_t} = U^t \ket{\psi_s} ~\text{for}~ t = 0, 1, 2, \dots.
	\end{equation}
	In terms of the density matrix, the state after $t$ steps is
	\begin{equation}\label{evolution_without_noise}
	\rho_t = \ket{\psi_t}\bra{\psi_t} = U^t \ket{\psi_s} \bra{\psi_s} (U^t)^\dagger.
	\end{equation} 
	
	In quantum walk, we say there is a perfect state transfer from the vertex $s$ to the vertex $r$ if there is a time $t = \tau$, such that $\ket{\psi_r} = \ket{\psi_\tau}$. In other words, the fidelity between $\ket{\psi_r}$ and $\ket{\psi_\tau}$ is $1$. Perfect state transfer is a rare phenomenon on simple graphs, particularly if it is related to a DTQW. There are graphs for which fidelity may not be equal to exactly $1$ \cite{kurzynski2011discrete, vstefavnak2016perfect, santos2022quantum, kendon2011perfect, vstefavnak2017perfect, yang2018quantum}. But, the value of fidelity may be high $> 0.8$. If $s = r$ and $\ket{\psi_\tau} = \ket{\psi_s}$ at time $t = \tau > 0$, we say that the walker is periodic with period $\tau$ at the vertex $s$. Periodicity and state transfer in quantum walk depend on the structure of the graphs. 
	
	\section{Quantum noise on quantum walk}
	
	We consider two quantum noises, which are the non-Markovian Random Telegraph Noise (RTN) and modified non-Markovian Ornstein-Uhlenbeck Noise (OUN). RTN exhibits oscillatory, damped behaviour in its non-Markovian regime, while OUN shows a power-law decay without oscillations \cite{utagi2020temporal}. These quantum noises were initially defined for qubit states. Due to the connection with quantum walk on graphs with an arbitrary number of vertices and edges, it is essential to generalize them to be applicable on quantum states in an arbitrary dimension. We use the Weyl operators $U_{u,v}$ for constructing quantum channels whose dimension is related to the order of the graphs under consideration.
	
	The non-Markovian RTN channel \cite{kumar2018non, banerjee2017non} can be represented by the Kraus operators 
	\begin{equation}\label{RTN}
	\begin{split}
	& K_1(t) = \sqrt{\frac{1+ \Lambda(t)}{2}}~ U_{0,0}, K_2(t) = \sqrt{\frac{1-\Lambda(t)}{2}}~ U_{1,0}, ~\text{where}\\
	& \Lambda(t) = e^{-\gamma t} \left[ \cos\left(\nu \gamma t \right) + \frac{\sin\left(\nu \gamma t \right)}{\nu} \right], ~\text{and} ~ \nu = {\sqrt{\left( \frac{2a}{\gamma} \right)^2 - 1}}.
	\end{split}
	\end{equation}	
	Here, $\nu$ is the frequency of the harmonic oscillators. The function $\Lambda(t)$ represents the damped harmonic function. We observe non-Markovian behaviour for $\frac{a}{\gamma} > 0.5$. For a fixed value of $a$ and $\gamma$, this channel depends on time $t$. Note that
	\begin{equation}
	\begin{split}
	K_{1}^\dagger K_{1} + K_{2}^\dagger K_{2}
	= & \left[\sqrt{\frac{1+ \Lambda(t)}{2}} U_{0,0}\right]^\dagger\left[\sqrt{\frac{1+ \Lambda(t)}{2}} U_{0,0}\right] + \left[\sqrt{\frac{1- \Lambda(t)}{2}} U_{1,0} \right]^\dagger \left[\sqrt{\frac{1+ \Lambda(t)}{2}} U_{1,0}\right]\\
	= & \left (\frac{1+ \Lambda(t)}{2}\right)U_{0,0}^\dagger U_{0,0} + \left (\frac{1+ \Lambda(t)}{2}\right) U_{1,0}^\dagger U_{1,0} = I_{d}.
	\end{split}
	\end{equation}
	Hence, these Kraus operators satisfy the completeness criterion. Here we take $a = 0.1$ and $\gamma = 0.01$, for all the considered graphs.
	
	Modified OUN also induces memory effects \cite{uhlenbeck1930theory, kumar2018non, banerjee2017non}. For the states of arbitrary dimension, this channel can be represented by the Kraus operators
	\begin{equation}\label{OUN}
	\begin{split}
	& K_1(t) = \sqrt{\frac{1+ P(t)}{2}}~ U_{0,0}, ~\text{and}~ K_2(t) = \sqrt{\frac{1- P(t)}{2}}~ U_{1,0}, ~\text{where}\\
	& P(t) = e^{-\frac{\lambda}{2} \left(t + \frac{1}{\gamma} \left(e^{-\gamma t} - 1\right)\right)}.
	\end{split}
	\end{equation}		
	Here, $\gamma $ specifies the noise bandwidth and $\lambda$ is the relaxation time. When $\gamma$ and $\lambda$ are constants, this channel depends on time $t$, only. Here we assume $\lambda = 1$ and $\gamma = 0.05$, for all the considered graphs. The completeness criterion follows from the equation below: 
	\begin{equation}
	\begin{split}
	K_{1}^\dagger K_{1} + K_{2}^\dagger K_{2} = & \left[\sqrt{\frac{1+ P(t)}{2}} U_{0,0}\right]^\dagger\left[\sqrt{\frac{1+ P(t)}{2}} U_{0,0}\right] + \left[\sqrt{\frac{1- P(t)}{2}} U_{1,0} \right]^\dagger \left[\sqrt{\frac{1+ P(t)}{2}} U_{1,0}\right]\\
	= & \left (\frac{1+ P(t)}{2}\right)U_{0,0}^\dagger U_{0,0} + \left (\frac{1+ P(t)}{2}\right) U_{1,0}^\dagger U_{1,0} = I_{d}.
	\end{split}
	\end{equation}
	The dimensions of the coin, shift and Kraus operators are $2m$ for all considered graphs.
	
	To apply the non-Markovian noises on quantum walk, we transfer the state $\rho_t$ via a noisy channel, where $t$ is finite. Both of the noises under our consideration are time-dependent. We consider the time parameter of noise equal to the time step of the quantum walker. This model of quantum noise has already been investigated in the literature \cite{kumar2018non, banerjee2017non, rani2024non}. Let the noise be represented by a set of Kraus operators $\{K_i\}$. After the $t$-th step of the quantum walk in the noisy environment, the state of the walker is represented by 
	\begin{equation}\label{evolution_with_noise}
	\rho'_t = \sum_i K_{i} \rho_t K_{i}^\dagger,
	\end{equation}
	where $\rho_t$ is mentioned in equation \eqref{evolution_without_noise}. If we do not consider the effect of noise in the system, the state of the walker is $\rho_t$, determined by equation \eqref{evolution_without_noise}. With noise, the state is $\rho'_t$ and is determined by equation \eqref{evolution_with_noise}.
	
	\section{State-transfer and periodicity on different graphs}
	
	The above construction is applicable to all simple graphs, in general. For our numerical analysis, we consider a number of graphs depicted in Figure \ref{simple_graphs}. We investigate state transfer and periodicity with and without noise on these graphs. These constructions can be extended to larger graphs. 
	
	\subsection{Quantum walk on path graphs}
	
	Consider the path graph $P_5$ depicted in subfigure \ref{fig:path_graph}. The directed graph $\overrightarrow{P_5}$ has $8$ edges which are $\overrightarrow{(0, 1)}, \overrightarrow{(1, 0)}, \overrightarrow{(1, 2)},$ $ \overrightarrow{(2, 1)}, \overrightarrow{(2, 3)}, \overrightarrow{(3, 2)}, \overrightarrow{(3, 4)}$, and $\overrightarrow{(4, 3)}$, arranged in a lexicographic order mentioned in equation (\ref{lexicographic_order}). 
	Here we observe two cases. 
	
	\begin{itemize}
		\item \textbf{Case $1$: Maximum fidelity is observed when the sender and receiver are located at the extreme points of the path graph.}\\
		
		\item \textbf{Case $2$: Minimum fidelity is observed when the sender and receiver are located at non-extreme (intermediate) positions on the path graph.}\\
		
	\end{itemize}
	In Case $1$, we assume a state transfer from $s = 0$ to $r = 4$. Therefore, using equation (\ref{sender_state}), we have the initial state of the walker
	\begin{equation}
	\ket{\psi_0} = (1, 0, 0, 0, 0, 0, 0, 0)^\dagger.
	\end{equation}
	To understand the state transfer, we calculate fidelity between $\ket{\psi_t} = U^t\ket{\psi_0}$ and 
	\begin{equation}
	\ket{\psi_4} = (0, 0, 0, 0, 0, 0, 0, 1)^\dagger.
	\end{equation}
	Note that, $\ket{\psi_4}$ is the receiver state mentioned in equation \eqref{receiver_state}. We denote the density matrix corresponding to $\ket{\psi_4}$ by $\rho_4$. The Grover diffusion operator for the extreme nodes as degree is one, is given as,
    \begin{equation}
        G_{0} = G_{4} = \begin{bmatrix}
            1
        \end{bmatrix}.
    \end{equation}

    Similarly, the Grover diffusion operator for non-extreme nodes with a degree two is expressed as follows,
    \begin{equation}\label{Grover_2}
        G_{1} = G_{2} = G_{3} = \begin{bmatrix}
            0 & 1 \\
            1 & 0 
        \end{bmatrix}.
    \end{equation}
	
	Using equations (\ref{Grover}) and (\ref{coin}), we derive the coin operators of the quantum walk, which we denote as  
	\begin{equation}\label{coin_state_transfer_P_5}
	C_{P_5} = \begin{bmatrix}
	-1 &  0 &  0 &  0 &  0 &  0 &  0 &  0\\
	0 & 0 &  1 &  0 &  0 &  0 &  0 &  0\\
	0 &  1 & 0 &  0 &  0 &  0 &  0 &  0\\
	0 &  0 &  0 & 0 &  1 &  0 &  0 &  0\\
	0 &  0 &  0 &  1 & 0 &  0 &  0 &  0\\
	0 &  0 &  0 &  0 &  0 & 0 &  1 &  0\\
	0 &  0 &  0 &  0 &  0 &  1 & 0 &  0\\
	0 &  0 &  0 &  0 &  0 &  0 &  0 &  -1
	\end{bmatrix}.
	\end{equation}
	Also, using equation \eqref{shift}, the shift operator can be represented by
	\begin{equation}
	S_{P_5} = \begin{bmatrix}
	0& 1 & 0 & 0& 0 & 0 & 0 & 0\\
	1 & 0& 0 & 0 & 0 & 0 & 0& 0\\
	0 & 0 & 0& 1 & 0 & 0 & 0 & 0\\
	0 & 0 & 1 & 0 & 0 & 0 & 0 & 0\\
	0 & 0 & 0 & 0 & 0 & 1 & 0 & 0\\
	0 & 0& 0 & 0 & 1 & 0 & 0 & 0\\
	0 & 0 & 0 & 0& 0 & 0 & 0 & 1\\
	0 & 0 & 0 & 0 & 0 & 0 & 1 & 0
	\end{bmatrix}.
	\end{equation}
	Therefore, the unitary operator of the DTQW is $U_{P_5} = S_{P_5} \times C_{P_5}$.
	
\begin{figure}
		\begin{subfigure}{1\textwidth}
			\includegraphics[width=\textwidth]{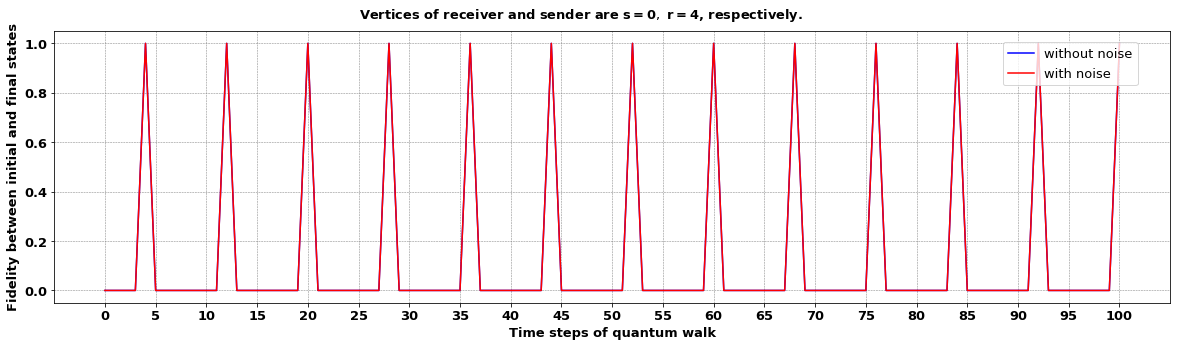}
			\caption{Fidelity with and without non-Markovian RTN noise when the sender and receiver are located at the extreme points.}
			\label{P_5_state_transfer_RTN} 
		\end{subfigure}
		\begin{subfigure}{1\textwidth}
			\includegraphics[width=\textwidth]{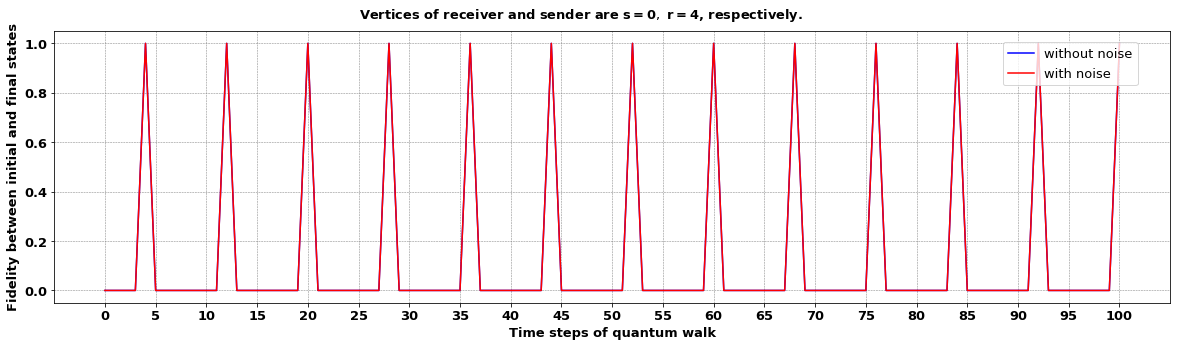}
			\caption{Fidelity with and without modified OUN noise when the sender and receiver are located at the extreme points.} 
			\label{P_5_state_transfer_OUN}
		\end{subfigure}
		\begin{subfigure}{1\textwidth}
			\includegraphics[width=\textwidth]{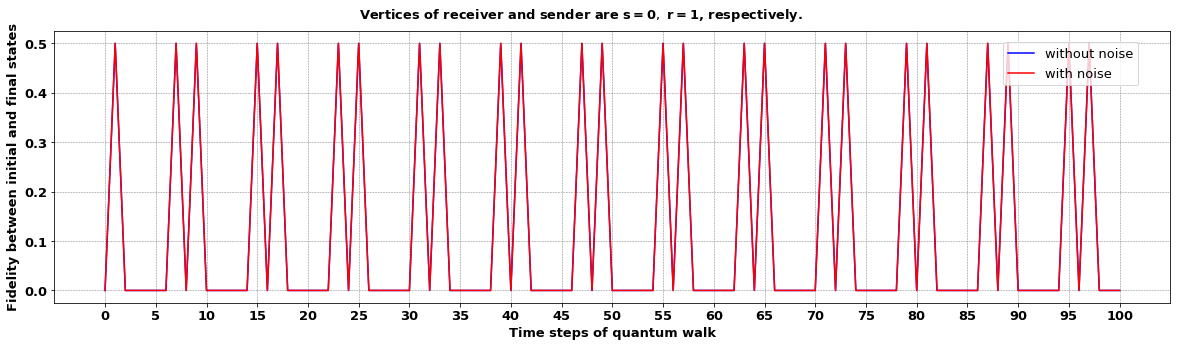}
			\caption{Fidelity with and without RTN noise when the sender and receiver are located at the intermediate points.}
			\label{P_5_state_transfer_(0,1)_RTN} 
		\end{subfigure}
		\begin{subfigure}{1\textwidth}
			\includegraphics[width=\textwidth]{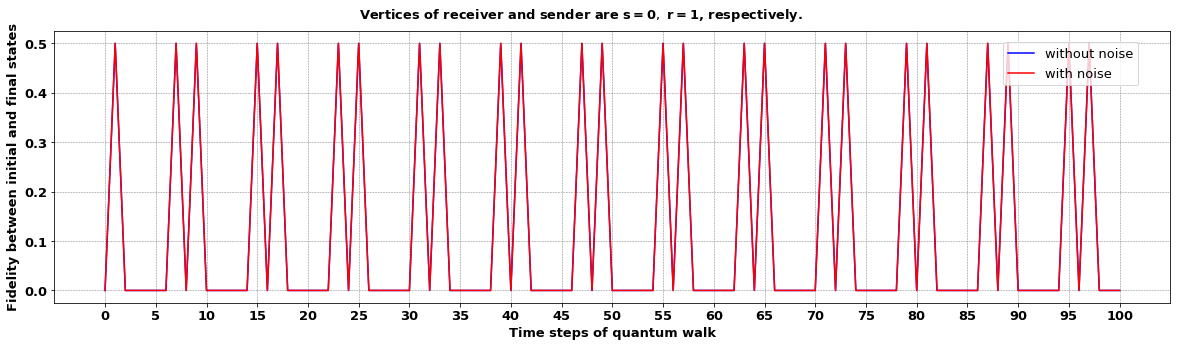}
			\caption{Fidelity with and without modified OUN noise when the sender and receiver are located at the intermediate points.} 
			\label{P_5_state_transfer_(0,1)_OUN}
		\end{subfigure}
		\caption{State-transfer on the path graph $P_{5}$ under non-Markovian RTN and modified non-Markovian OUN noise. The RTN parameters are set to $a = 0.1$, $\gamma = 0.01$, while the OUN parameters are $\lambda = 1$, $\gamma = 0.05$.}
		\label{P5_fidelity_state_transfer}
	\end{figure}

	Now, we apply the operator $U_{P_5}$ on the sender state $\ket{\psi_0}$, to find $\ket{\psi_t}$ for different time steps. We compute $F(\ket{\psi_t}, \ket{\psi_4})$ and $F(\ket{\rho'_t}, \ket{\rho_4})$, which is the fidelity without noise and with noise using equations \eqref{fidelity_vectors} and \eqref{fidelity_density}, respectively. The variation in fidelity under RTN and OUN noise and without noise with time steps, for Case $1$, is shown in the subfigures \ref{P_5_state_transfer_RTN} and \ref{P_5_state_transfer_OUN}, respectively. It can be verified from these subfigures that the walker moves from vertex $0$ to vertex $4$ at time steps $t = 4, 12, 20, \dots$ under noise and without noise. 
	Therefore, we can conclude that there is no effect of OUN and RTN noise on state transfer on the path graph with time steps. Similarly, for Case $2$, we take $s=0, r=1$ and observe that the fidelity becomes half with and without noise, as depicted in subfigures \ref{P_5_state_transfer_(0,1)_RTN} and \ref{P_5_state_transfer_(0,1)_OUN}.
	
\begin{figure}
		\begin{subfigure}{1\textwidth}
			\includegraphics[width=\textwidth, height = 4.5cm]{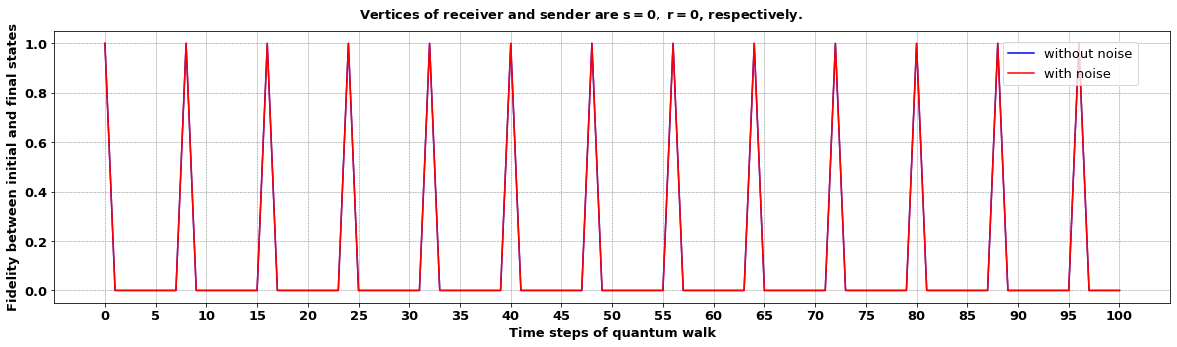}
			\caption{Fidelity with and without RTN noise when both the sender and receiver are located at the central vertex.} 
			\label{P_5_periodicity_RTN}
		\end{subfigure}
		\begin{subfigure}{1\textwidth}
			\includegraphics[width=\textwidth, height = 4.5cm]{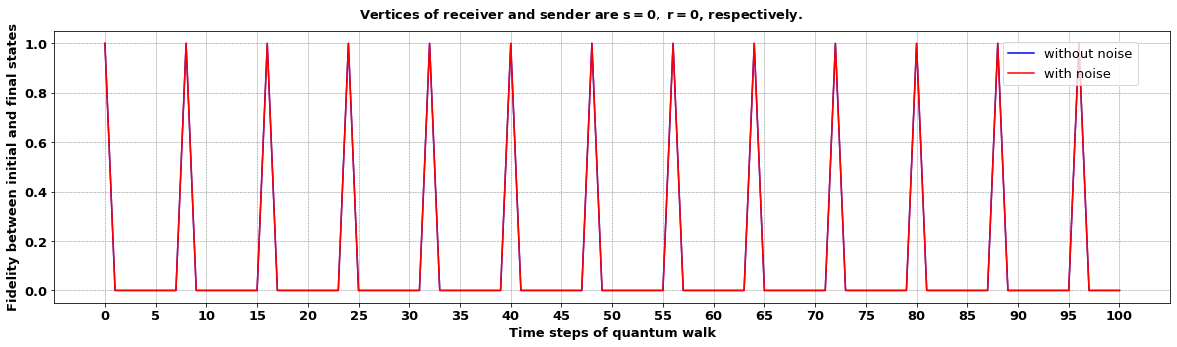}
			\caption{Fidelity with and without OUN noise when both the sender and receiver are located at the central vertex.}
			\label{P_5_periodicity_OUN}
		\end{subfigure}
		\caption{Periodicity on the path graph $P_{5}$ under non-Markovian RTN and modified non-Markovian OUN noise. The RTN parameters are set to $a = 0.1$, $\gamma = 0.01$, while the OUN parameters are $\lambda = 1$, $\gamma = 0.05$.}
		\label{P5_fidelity_periodicity}
	\end{figure}

	Further, we study the periodicity of the quantum walker on the path graph at vertex $0$ with and without noise. Assuming $s = r = 0$, we update the coin operators as 
	\begin{equation}
	C'_{P_5} = \begin{bmatrix}
	-1 & 0 & 0 & 0 & 0 & 0 & 0 & 0\\
	0 & 0 & 1 & 0 & 0 & 0 & 0 & 0\\
	0 & 1 & 0 & 0 & 0 & 0 & 0 & 0\\
	0 & 0 & 0 & 0 & 1 & 0 & 0 & 0\\
	0 & 0 & 0 & 1 & 0 & 0 & 0 & 0\\
	0 & 0 & 0 & 0 & 0 & 0 & 1 & 0\\
	0 & 0 & 0 & 0 & 0 & 1 & 0 & 0\\
	0 & 0 & 0 & 0 & 0 & 0 & 0 & 1
	\end{bmatrix},
	\end{equation} 
	which is different from $C_{P_5}$ in equation \eqref{coin_state_transfer_P_5}. But, there is no change in the shift operator. Any update in the coin operator modifies the unitary operator, which is now $U'_{P_5} = S_{P_5} \times C'_{P_5}$. Now, we again compute the fidelity under non-Markovian RTN and OUN noise and observe the periodicity with time steps. From the subfigures \ref{P_5_periodicity_RTN} and \ref{P_5_periodicity_OUN}, it can be verified that the walker returns to vertex $0$ periodically at time $t = 0, 8, 16, \dots $. Also, the RTN and OUN noise do not affect periodicity on the path graph with time steps.
	

    We can observe from the figures that state transfer fidelity is maximized when the sender and receiver are at the end vertices of the path graph. In our model, the noise is coupled to the degree of the vertices. At the ends of a path graph, every vertex has degree $1$. Therefore, the quantum walker has only one available direction to move. This prevents amplitude splitting and preserves the chances for constructive interference under noise, thereby maximizing the state transfer fidelity. At an interior vertices, the walkers can move left or right. Hence, the amplitude splits into multiple interfering paths. As a result, the fidelity of state transfer drops due to destructive interference.
    
	\subsection{Quantum walk on cycle graphs}
	
	In this subsection, we consider the cycle graph $C_6$ with $6$ vertices, depicted in subfigure \ref{fig:cycle_graph}, to investigate state transfer and periodicity. We observe two cases.
	\begin{itemize}
		\item \textbf{Case $1$: The sender and receiver are at the opposite vertices of the cycle graph.}\\
		
		\item \textbf{Case $2$: The sender and receiver are asymmetrically placed on the cycle graph.}\\
	\end{itemize}
	
	For Case $1$, we apply $ s=0$ and $r=3$. Following equation \eqref{sender_state} we have			
	\begin{equation}
	\ket{\psi_0} = \frac{1}{\sqrt{2}}(1, 1, 0, 0, 0, 0, 0, 0, 0, 0, 0, 0)^\dagger.
	\end{equation}
	To understand state transfer, we calculate fidelity between $\ket{\psi_t}$ and 
	\begin{equation}
	\ket{\psi_3} = \frac{1}{\sqrt{2}}(0, 0, 0, 0, 0, 0, 1, 1, 0, 0, 0, 0)^\dagger.
	\end{equation}

    The Grover diffusion operator for all the nodes with a degree two is as in equation (\ref{Grover_2}).

	Using equations (\ref{Grover}) and (\ref{coin}), we derive the coin operators for the quantum walk, which is 
	\begin{equation}\label{coin_state_transfer_C_6}
	C_{C_6} = \begin{bmatrix}
	0 & -1 &  0 &  0 & 0 &  0 &  0 &  0 &  0 & 0 &  0 &  0\\
	-1 &  0 &  0 &  0 &  0 & 0 &  0 &  0 & 0 &  0 &  0 & 0\\
	0 &  0 & 0 &  1 &  0 &  0 &  0 &  0 &  0 &  0 & 0 &  0\\
	0 &  0 &  1 & 0 & 0 &  0 &  0 &  0 & 0 &  0 &  0 &  0\\
	0 & 0 &  0 & 0 & 0 & 1 &  0 &  0 & 0 & 0 & 0 &  0\\
	0 & 0 & 0 &  0 &  1 & 0 &  0 & 0 & 0 & 0 & 0 &  0\\
	0 & 0 & 0 & 0 & 0 & 0 & 0 & -1 & 0 &  0 & 0 & 0 \\
	0 &  0& 0 & 0 & 0 & 0 & -1 & 0 & 0 &  0 &  0 &  0\\
	0 & 0 & 0 & 0 &  0 & 0 & 0 & 0 & 0 & 1 & 0 & 0 \\
	0 & 0 & 0 & 0 &  0 & 0 & 0 & 0 & 1 & 0 & 0 &  0\\
	0 & 0 & 0 & 0 &  0 &  0 & 0 & 0 & 0 & 0 & 0 & 1\\
	0 & 0 & 0 & 0 & 0 &  0 &  0 & 0 &  0 & 0 & 1 & 0
	\end{bmatrix}.
	\end{equation}
	Also, using equation \eqref{shift}, the shift operator can be represented by
	\begin{equation}
	S_{C_6} = \begin{bmatrix}
	0 & 0 & 1 &0 & 0 & 0 & 0 & 0 & 0 & 0 & 0 & 0\\
	0 & 0 & 0 & 0 & 0 & 0 & 0& 0 & 0 & 0 & 1 & 0\\
	1 & 0 & 0 & 0 & 0 & 0 & 0 & 0 & 0 & 0 & 0 & 0\\
	0 & 0 & 0 & 0 & 1 & 0 & 0 & 0 & 0 & 0 & 0 & 0\\
	0 & 0 & 0 & 1 & 0 & 0 & 0 & 0 & 0 & 0 & 0 & 0\\
	0 & 0 & 0 & 0 & 0 & 0 & 1 & 0 & 0 & 0 & 0 & 0\\
	0 & 0 & 0 & 0 & 0 & 1 & 0 & 0 & 0 & 0 & 0 & 0\\
	0 & 0 & 0 & 0 & 0 & 0 & 0 & 0 & 1 & 0 & 0 & 0\\
	0 & 0 & 0 & 0 & 0 & 0 & 0 & 1 & 0 & 0 & 0 & 0\\
	0 & 0 & 0 & 0 & 0 & 0 & 0 & 0 & 0 & 0 & 0 & 1\\
	0 & 1 & 0 & 0 & 0 & 0 & 0 & 0 & 0 & 0 & 0 & 0\\
	0 & 0 & 0& 0 & 0 & 0 & 0 & 0 & 0 & 1 & 0 & 0
	\end{bmatrix}.
	\end{equation}
	Therefore, the unitary operator of the DTQW is $U_{C_6} = S_{C_6} \times C_{C_6}$. Next, we apply the operator $U_{C_6}$ on the sender state $\ket{\psi_0}$, to compute $\ket{\psi_t}$ for different time steps. 

    In this article, we consider the cycle graphs with an even number of vertices only. In a cycle graph, every vertex has two neighbours. When the sender and the receiver are placed opposite to each other on an even cycle, maintaining the topology of a cycle graph we can construct two paths of equal length from the sender to the receiver. These two paths leads to clockwise and counterclockwise movement on the cycle. The symmetry of the cycle graph ensures that when the sender and receiver are placed opposite to each other, quantum amplitudes travelling clockwise and counterclockwise accumulate at identical phases. It leads to constructive interference that preserves fidelity. This constructive interference effect persists across all even cycles. 
    
	This sheds light into the state transfer in a noisy environment on a cycle graph for Case $1$. Here, the sender reaches the receiver vertex $3$ via two symmetric paths 0 $\rightarrow 1 \rightarrow 2 \rightarrow 3 \quad \text{and} \quad 0 \rightarrow 5 \rightarrow 4 \rightarrow 3$. Hence, the sender and receiver are separated by an even number of steps. The graph symmetry allows two equal-length paths (clockwise and counterclockwise) leading to constructive interference.   It can be observed that the non-Markovian RTN and OUN noises preserve this constructive interference, as depicted in the subfigures \ref{C_6_state_transfer_(0,3)_RTN} and \ref{C_6_state_transfer_(0,3)_OUN}. This holds for any two opposite sender and receiver positions in a cycle graph. 
	
\begin{figure}
		\begin{subfigure}{1\textwidth}
			\includegraphics[width=\textwidth, height = 4.5cm]{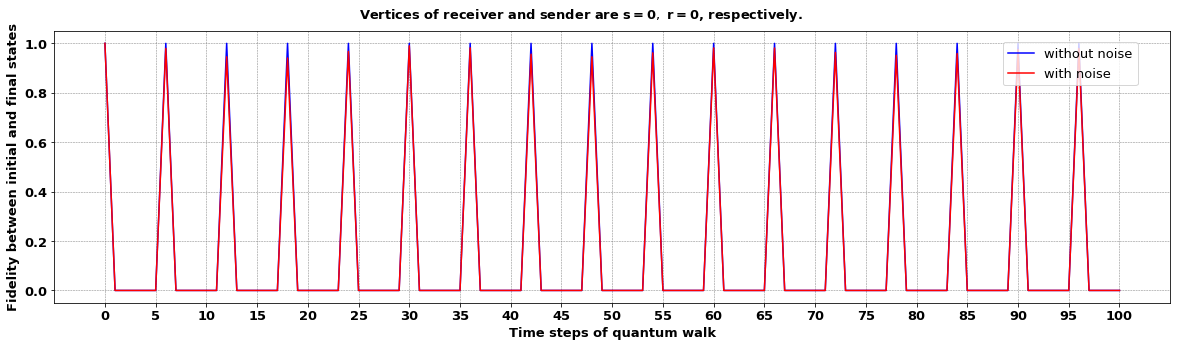}
			\caption{Fidelity with and without RTN noise when both the sender and receiver are located at the central vertex.} 
			\label{C_6_periodicity_RTN}
		\end{subfigure}
		\begin{subfigure}{1\textwidth}
			\includegraphics[width=\textwidth, height = 4.5cm]{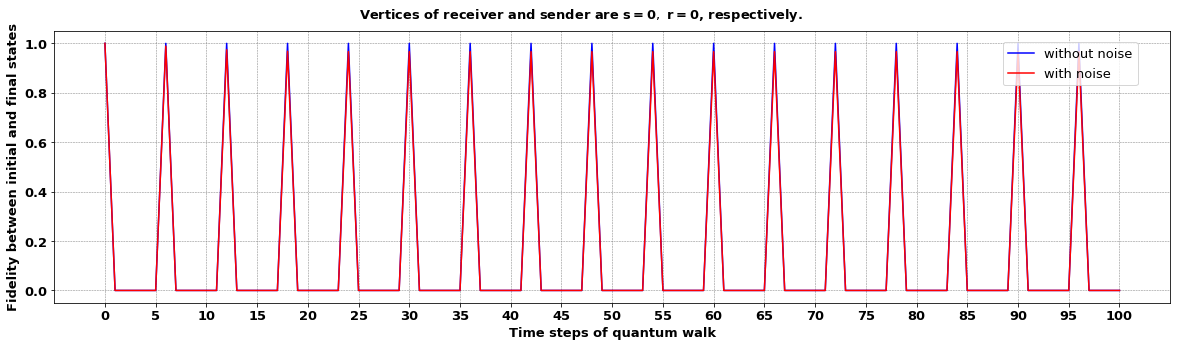}
			\caption{Fidelity with and without OUN noise when both the sender and receiver are located at the central vertex.}
			\label{C_6_periodicity_OUN}
		\end{subfigure}
		\caption{Periodicity at vertex $0$ on the cycle graph $C_{6}$ under non-Markovian RTN and non-Markovian OUN noise. The channel parameters are the same as for the path graph.}
		\label{C6_fidelity_periodicity}
	\end{figure}

	In Case $2$, we assume that the sender is at vertex $0$ and the receiver is at vertex $1$. Here, the sender and the receiver are separated by an odd or unequal number of steps. In this scenario, the paths connecting them are not symmetric, and destructive interference with and without noise results in a sharp drop in fidelity, as depicted in the subfigures \ref{C_6_state_transfer_(0,1)_RTN} and \ref{C_6_state_transfer_(0,1)_OUN}. Further, we observe the periodicity of the quantum walk on a cycle graph putting $s = r = 0$. It is seen that there is a minor effect of noise on the periodicity of the cycle graph. Subfigures \ref{C_6_periodicity_RTN} and \ref{C_6_periodicity_OUN} present the comparison between fidelity values with and without noise. 
	\begin{figure}
		\begin{subfigure}{1\textwidth}
			\includegraphics[width=\textwidth]{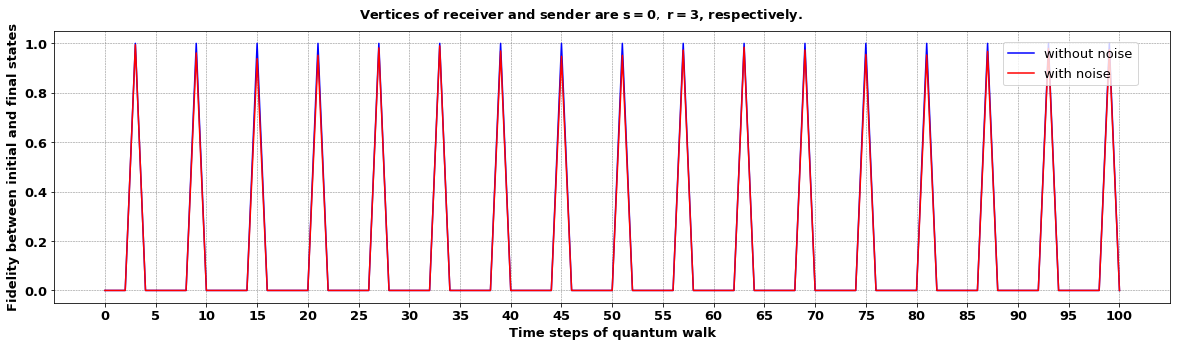}
			\caption{Fidelity with and without RTN noise when sender and receiver are at the opposite vertices of the cycle graph.}
			\label{C_6_state_transfer_(0,3)_RTN} 
		\end{subfigure}
		\begin{subfigure}{1\textwidth}
			\includegraphics[width=\textwidth]{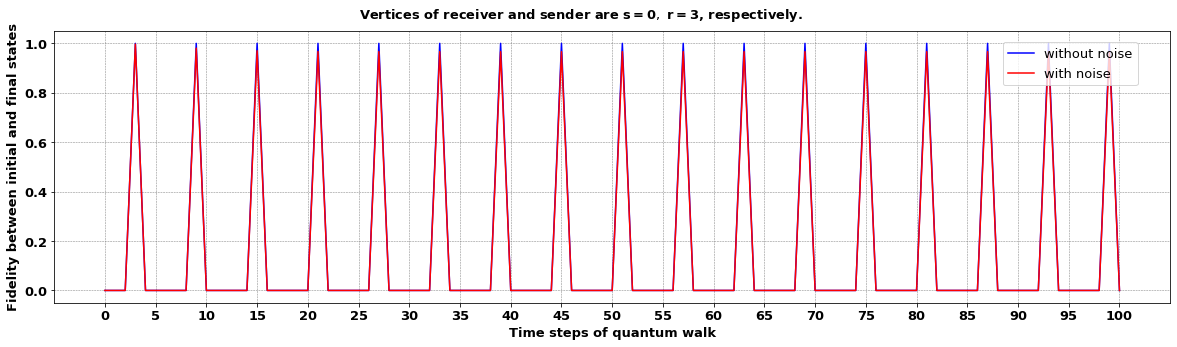}
			\caption{Fidelity with and without OUN noise when the sender and receiver are placed at opposite vertices of the cycle graph.} 
			\label{C_6_state_transfer_(0,3)_OUN}
		\end{subfigure}
		\begin{subfigure}{1\textwidth}
			\includegraphics[width=\textwidth]{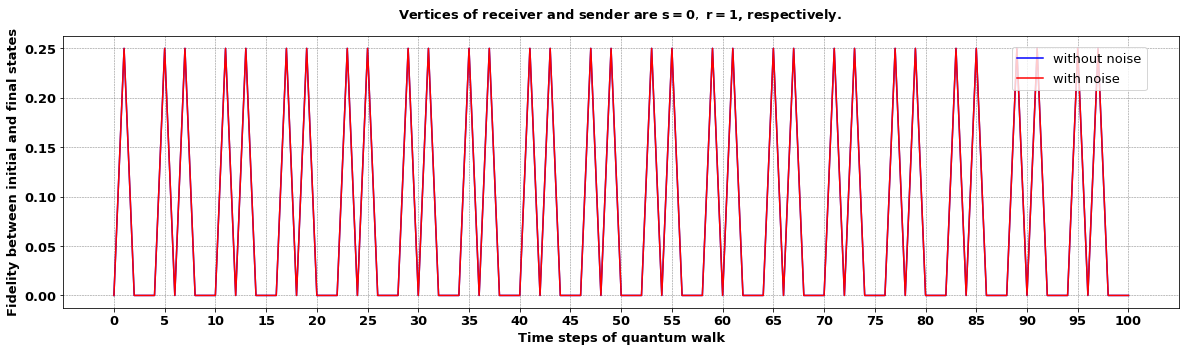}
			\caption{Fidelity with and without RTN noise when the sender and receiver are placed asymmetrically on the cycle graph.}
			\label{C_6_state_transfer_(0,1)_RTN} 
		\end{subfigure}
		\begin{subfigure}{1\textwidth}
			\includegraphics[width=\textwidth]{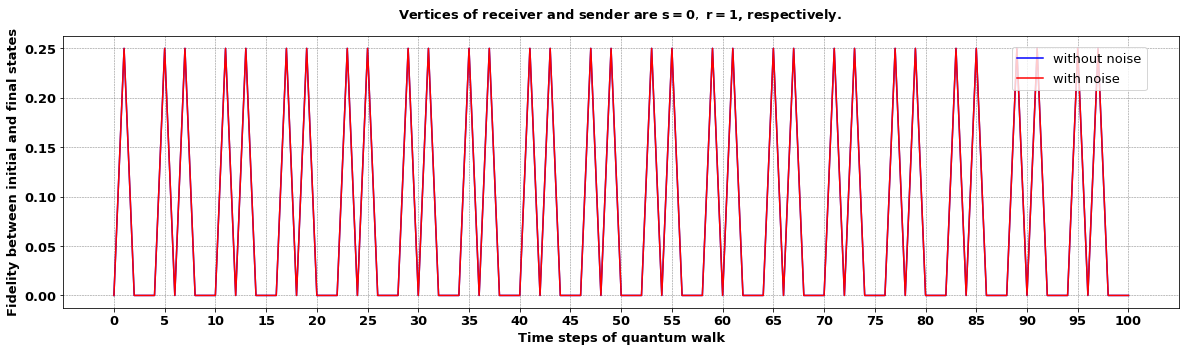}
			\caption{Fidelity with and without OUN noise when the sender and receiver are placed asymmetrically on the cycle graph.} 
			\label{C_6_state_transfer_(0,1)_OUN}
		\end{subfigure}
		\caption{State transfer on the cycle graph $C_{6}$ under non-Markovian RTN and non-Markovian OUN noise. The channel parameters are the same as for the path graph.}
		\label{C6_fidelity_state_transfer}
	\end{figure}
	
	\subsection{Quantum walk on star graphs}
	
	Now we describe state transfer and periodicity on the star graph $S_{6}$, depicted in sub-figure \ref{fig:star_graph}. We investigate two cases.
	\begin{itemize}
		\item \textbf{Case 1: Sender at the central node and receiver at any external node.}\\
		\item \textbf{Case 2: Sender at any external node and receiver at the central node.}\\
	\end{itemize}
	
	We observe that in Case $1$, there is no effect of noise on the fidelity of the state transfer on the star graph with time steps. We consider the sender $s = 0$ and receiver $r = 1$. Therefore, the initial state is
	\begin{equation}
	\ket{\psi_0} = \frac{1}{\sqrt{5}}(1, 1, 1, 1, 1, 0, 0, 0, 0, 0)^\dagger,
	\end{equation}
	and the receiver state is 
	\begin{equation}
	\ket{\psi_1} = (0, 0, 0, 0, 0, 1, 0, 0, 0, 0)^\dagger.
	\end{equation}

        The Grover diffusion operator for the central node with degree five is given as,
    \begin{equation}
        G_{0} = G_{4} = \begin{bmatrix}
            -0.6 & 0.4 & 0.4 & 0.4 & 0.4 \\
             0.4 & -0.6 & 0.4 & 0.4 & 0.4 \\
             0.4 & 0.4 & -0.6 & 0.4 & 0.4 \\
             0.4 & 0.4 & 0.4 & -0.6 & 0.4 \\
             0.4 & 0.4 & 0.4 & 0.4 & -0.6
        \end{bmatrix}.
    \end{equation}

    Similarly, the Grover diffusion operator for the external nodes with a degree one is expressed as follows,
    \begin{equation}
        G_{1} = G_{2} = G_{3} = G_{4} = G_{5} = \begin{bmatrix}
            1
        \end{bmatrix}.
    \end{equation}
	Using equations (\ref{Grover}) and (\ref{coin}), we derive the coin operators for the quantum walk for the star graph $S_{6}$, which is 
	\begin{equation}\label{coin_state_transfer_S_6}
	C_{S_6} = \begin{bmatrix}
	0.6 & -0.4 & -0.4 & -0.4 & -0.4 & 0 & 0 & 0 & 0 & 0\\
	-0.4 & 0.6 & -0.4 & -0.4 & -0.4 & 0 & 0 & 0 & 0 & 0\\
	-0.4 & -0.4 & 0.6 & -0.4 & -0.4 & 0 &  0 & 0 & 0 & 0\\
	-0.4 & -0.4 & -0.4 & 0.6 & -0.4 & 0 & 0 & 0 & 0 & 0\\
	-0.4 & -0.4 & -0.4 & -0.4 & 0.6 & 0 & 0 & 0 & 0 & 0\\
	0 & 0 & 0 & 0 & 0 & -1 &  0 & 0 &  0 & 0\\
	0 &  0 & 0 & 0 &  0 & 0 & 1 &  0 & 0 & 0\\
	0 & 0 &  0 &  0 &  0 & 0 &  0 & 1 & 0 & 0\\
	0 &0 & 0 & 0 & 0 & 0 & 0 & 0 & 1 &  0\\
	0 &  0 & 0 &  0 &  0 &  0 & 0 & 0 &  0 &  1
	\end{bmatrix}.
	\end{equation}
	Also using equation \eqref{shift}, the shift operator for the star graph $S_{6}$ can be represented by
	\begin{equation}
	S_{S_6} = \begin{bmatrix}
	0 & 0 & 0 & 0 & 0 & 1 & 0 & 0 & 0 & 0\\
	0 & 0 & 0 & 0 & 0 & 0 & 1 & 0 & 0 & 0\\
	0 & 0 & 0 & 0 & 0 & 0 & 0 & 1 & 0 & 0\\
	0 & 0 &0 & 0 & 0 & 0 & 0 & 0 & 1 & 0\\
	0 & 0 & 0 & 0 & 0 & 0 & 0 & 0 & 0 & 1\\
	1 & 0 & 0 & 0 & 0 & 0 & 0 & 0 & 0 & 0 \\
	0 & 1 & 0 & 0 & 0 & 0 & 0 & 0 & 0 & 0\\
	0 & 0 & 1 & 0 & 0 & 0 & 0 & 0 & 0 & 0\\
	0 & 0 & 0 & 1 & 0 & 0 & 0 & 0 &0 & 0\\
	0 & 0 & 0 & 0 & 1 & 0 & 0 & 0 & 0 & 0
	\end{bmatrix}.
	\end{equation}
	Therefore, the unitary operator of the DTQW is $U_{S_6} = S_{S_6} \times C_{S_6}$.
	
	We calculate fidelity $F(\ket{\psi_t}, \ket{\psi_1})$ without noise and $F(\rho'_t, \rho_1)$ with RTN and OUN noises. 
	It is observed that in Case $1$ ($s=0, r=1$), there is no effect of both non-Markovian RTN and OUN noises on the state transfer on the star graph shown in the sub-figures \ref{S_6_state_transfer_(0,1)_RTN} and \ref{S_6_state_transfer_(0,1)_OUN}. In this scenario, the quantum walker is initially localised at the central hub of the star graph. Here, the central node is directly connected to all external nodes. As a result, due to the strong central connectivity and equal probability of moving along any edge connected to the sender vertex, the noise has a similar impact on all paths; therefore, the fidelity under noise and without noise overlap. While in the Case $2$ ($s=1, r=0$), the walker can go only via the centre, and therefore the evolution becomes highly sensitive to noise, resulting in decrease in fidelity with noise, as shown in subfigures \ref{S_6_state_transfer_(1,0)_RTN} and \ref{S_6_state_transfer_(1,0)_OUN}.
	
\begin{figure}
		\begin{subfigure}{1\textwidth}
			\includegraphics[width=\textwidth, height = 4.5cm]{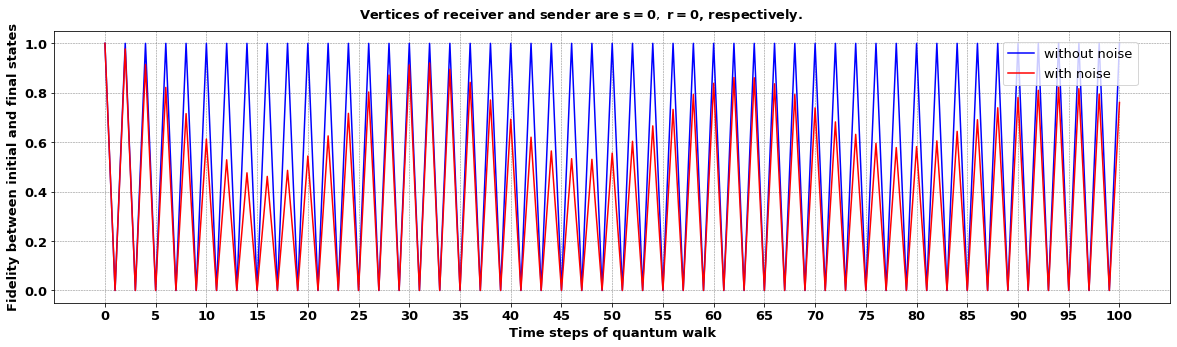}
			\caption{Fidelity with and without RTN noise when both the sender and receiver are located at the central vertex.} 
			\label{S_6_periodicity_RTN}
		\end{subfigure}
		\begin{subfigure}{1\textwidth}
			\includegraphics[width=\textwidth, height = 4.5cm]{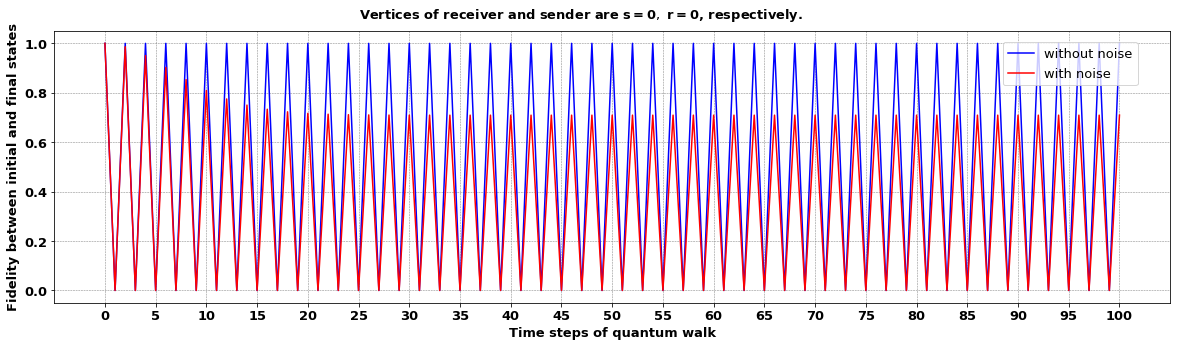}
			\caption{Fidelity with and without OUN noise when both the sender and receiver are located at the central vertex.}
			\label{S_6_periodicity_OUN}
		\end{subfigure}
		\caption{Periodicity at vertex $0$ on the star graph $S_{6}$ under non-Markovian RTN and modified non-Markovian OUN noise. The channel parameters are the same as for the path graph.}
		\label{S6_fidelity_periodicity}
	\end{figure}

	Now, we study the periodicity of the quantum walker at vertex $0$, with and without noise. We consider $s = r = 0$. The coin operators is updated as 
	\begin{equation}
	C'_{S_6} = \begin{bmatrix}
	0.6 & -0.4 & -0.4 & -0.4 & -0.4 & 0 & 0 & 0 & 0 & 0\\
	-0.4 & 0.6 & -0.4 & -0.4 & -0.4 & 0 & 0 & 0 & 0 & 0\\
	-0.4 & -0.4 & 0.6 & -0.4 & -0.4 & 0 &  0 & 0 & 0 & 0\\
	-0.4 & -0.4 & -0.4 & 0.6 & -0.4 & 0 & 0 & 0 & 0 & 0\\
	-0.4 & -0.4 & -0.4 & -0.4 & 0.6 & 0 & 0 & 0 & 0 & 0\\
	0 & 0 & 0 & 0 & 0 & -1 &  0 & 0 &  0 & 0\\
	0 &  0 & 0 & 0 &  0 & 0 & 1 &  0 & 0 & 0\\
	0 & 0 &  0 &  0 &  0 & 0 &  0 & 1 & 0 & 0\\
	0 &0 & 0 & 0 & 0 & 0 & 0 & 0 & 1 &  0\\
	0 &  0 & 0 &  0 &  0 &  0 & 0 & 0 &  0 &  1
	\end{bmatrix},
	\end{equation} 
	which is different from $C_{S_6}$ in equation \eqref{coin_state_transfer_S_6}.
	The sub-figures \ref{S_6_periodicity_RTN} and \ref{S_6_periodicity_OUN} show the comparison of the periodic behaviour of the walker without and with noise. From these sub-figures, we can observe that there are perfect periodic revivals without noise, but under non-Markovian RTN, the fidelity starts decreasing after time-step $2$, albeit with subsequent revivals and with a periodic, oscillatory envelope. This is a signature of information flowing back to the system from the environment and causes revivals in RTN. However, under non-Markovian OUN, the fidelity is decreasing after a few time steps and does not exhibit any information backflow \cite{utagi2020temporal, kumar2018non}. 
	\begin{figure}
		\begin{subfigure}{1\textwidth}\end{subfigure}
		\begin{subfigure}{1\textwidth}
			\includegraphics[width=\textwidth]{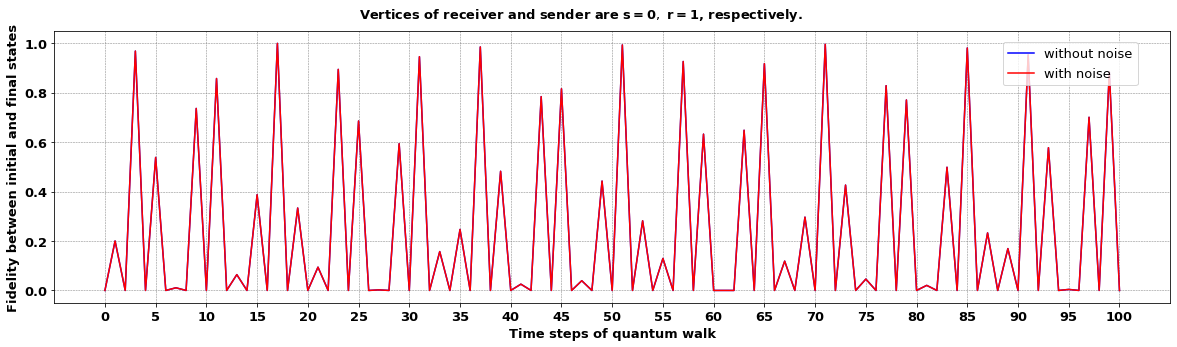}
			\caption{Fidelity with and without RTN noise when sender and receiver are placed at the central and external vertices, respectively.}
			\label{S_6_state_transfer_(0,1)_RTN} 
		\end{subfigure}
		\begin{subfigure}{1\textwidth}\end{subfigure}
		\begin{subfigure}{1\textwidth}
			\includegraphics[width=\textwidth]{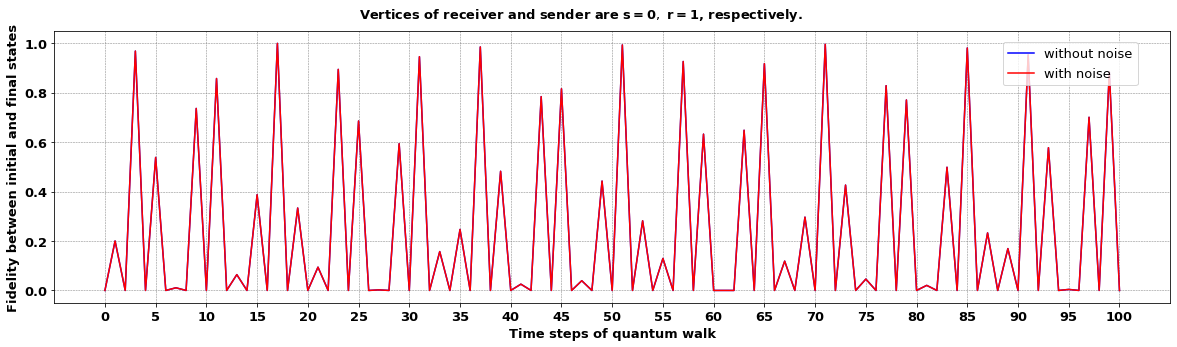}
			\caption{Fidelity with and without OUN noise when sender and receiver are placed at the central and external vertices, respectively.} 
			\label{S_6_state_transfer_(0,1)_OUN}
		\end{subfigure}
		\begin{subfigure}{1\textwidth}
			\includegraphics[width=\textwidth]{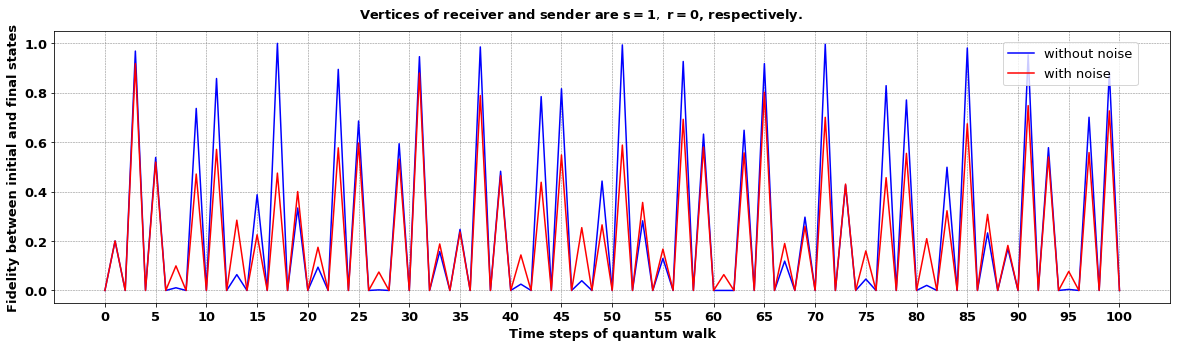}
			\caption{Fidelity with and without RTN noise when sender and receiver are placed at the external and central vertices, respectively.}
			\label{S_6_state_transfer_(1,0)_RTN} 
		\end{subfigure}
		\begin{subfigure}{1\textwidth}
			\includegraphics[width=\textwidth]{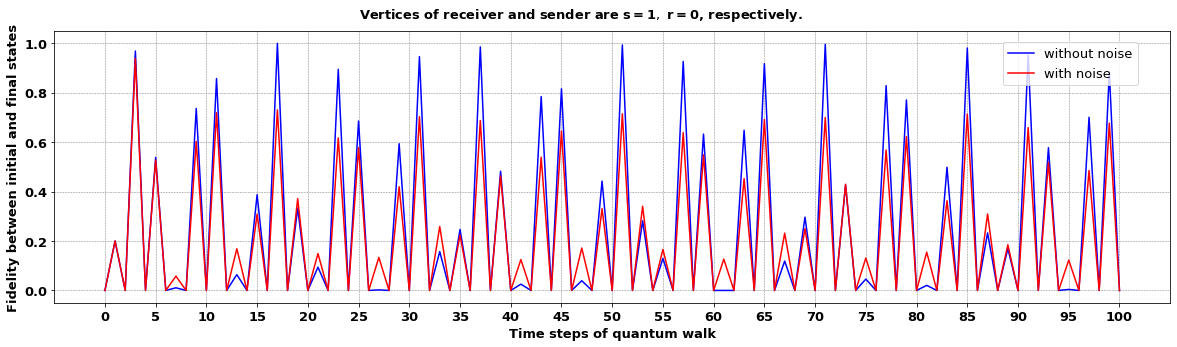}
			\caption{Fidelity with and without OUN noise when sender and receiver are placed at the external and central vertices, respectively.} 
			\label{S_6_state_transfer_(1,0)_OUN}
		\end{subfigure}
		\caption{State-transfer on the star graph $S_{6}$ under non-Markovian RTN and non-Markovian OUN noise. The channel parameters are the same as for the path graph.}
		\label{S6_fidelity_state_transfer}
	\end{figure}	

    The fidelity of quantum state transfer and periodicity of quantum walks are influenced by the channel parameters governing the non-Markovian noise models. For the star graph \ref{fig:star_graph}, we observe a prominent effect of noise parameters.
    
\begin{figure}
	\begin{subfigure}{0.45\textwidth}
		\includegraphics[scale = .5]{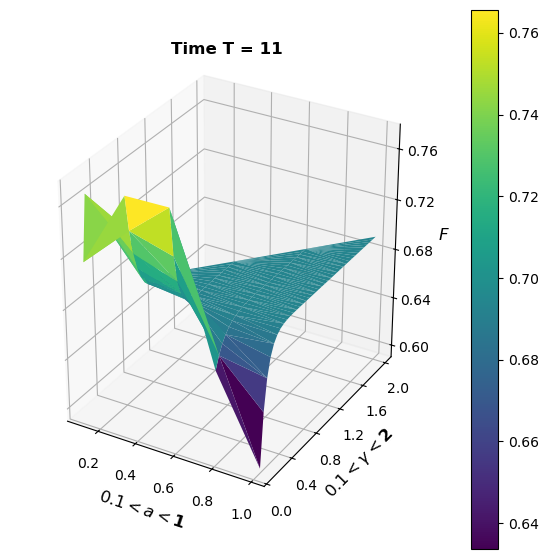}
            \caption{Fidelity with RTN noise when the state is transferred from an external vertex $s = 1$ to the central vertex $r = 0$.}
		\label{S6_state_transfer}
	\end{subfigure}
        \hspace{.5cm}
        \begin{subfigure}{0.45\textwidth}
		\includegraphics[scale = .5]{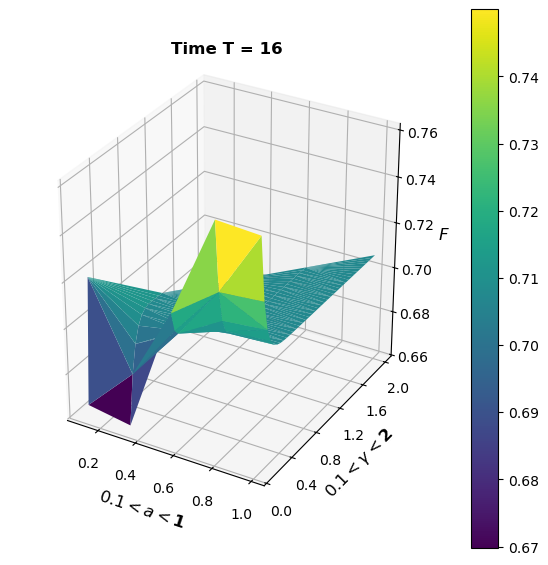}
		\caption{Fidelity with RTN noise when both the sender and receiver are located at the central vertex.} 
		\label{S6_periodiocity}
		\end{subfigure}
		\caption{Fidelity of state transfer and periodicity with varying channel parameters under non-Markovian RTN for the star graph \ref{fig:star_graph}.}
		\label{}
    \end{figure}

    In case of RTN noise, we consider 50 different values of $a$ with $0.1 < a < 5$. For every value of $a$, the other variable $\gamma$ varies from $0.1$ to $2a$. Let $s = 1$ and $r = 0$. At time $t = 11$, the surface plot of Fidelity is depicted in sub-figure \ref{S6_state_transfer}. 
    We also consider the periodicity of the quantum walk at $s = 0$ as depicted in sub-figure \ref{S6_periodiocity}. For the above mentioned range of $a$ and $\gamma$, we plot the Fidelity at $t = 16$. In both cases, we observe the periodic nature of fidelity.

	\subsection{Quantum walk on complete bipartite graphs}
	Now, we consider the complete bipartite graph $K_{2,3}$ for our investigation, which is depicted in sub-figure \ref{fig:complete_bi_graph}. Here, we observe the state transfer between the vertices $s=0$ and $r=1$. Hence, the initial state of the sender is
	\begin{equation}
	\ket{\psi_0} = \frac{1}{\sqrt{3}}(0, 0, 0, 1, 1, 1, 0, 0, 0, 0, 0,0)^\dagger,
	\end{equation}
	and the receiver state is
	\begin{equation}
	\ket{\psi_1} = \frac{1}{\sqrt{3}}(1, 1, 1, 0, 0, 0, 0, 0, 0, 0, 0, 0)^\dagger.
	\end{equation}

        The Grover diffusion operator for the first partition set with degree three is given as,
    \begin{equation}
        G_{0} = G_{1} = \begin{bmatrix}
            -0.33 & 0.67 & 0.67 \\
             0.67 & -0.33 & 0.67 \\
             0.67 & 0.67 & -0.33 
        \end{bmatrix}.
    \end{equation}
    Similarly, the Grover diffusion operator for the second partition set with a degree two is as in equation (\ref{Grover_2}).
    
	Using equations (\ref{Grover}) and (\ref{coin}), we derive the coin operators for the quantum walk for the complete bipartite graph $K_{(2,3)}$, which is 
	\begin{equation}\label{coin_state_transfer_K_{(2,3)}}
	C_{K_{(2,3)}} = \begin{bmatrix}
	0.33 & -0.67 & -0.67 & 0 &  0 & 0 & 0 & 0 & 0 & 0 & 0 & 0\\
	-0.67 & 0.33 & -0.67 &  0 & 0 & 0 & 0 & 0 &  0& 0 & 0 & 0\\
	-0.67 & -0.67 & 0.33 & 0 & 0 & 0 & 0 & 0 & 0 & 0 & 0 & 0\\
	0 & 0 & 0 & 0.33 &-0.67 & -0.67 & 0 & 0 & 0 & 0 & 0 & 0\\
	0 & 0 & 0 & -0.67 & 0.33 & -0.67 & 0 & 0 & 0 & 0 & 0 & 0\\
	0 & 0 & 0 & -0.67 & -0.67 & 0.33 & 0 & 0 & 0 & 0 & 0 &  0\\
	0 & 0 & 0 & 0 & 0 & 0 & 0 & 1 & 0 & 0 & 0 & 0\\
	0 &  0 & 0 & 0 & 0 & 0 & 1 & 0 & 0 & 0 & 0 & 0 \\
	0 & 0 & 0 & 0 & 0 & 0 & 0 & 0 & 0 & 1 & 0 & 0 \\
	0 & 0 & 0 & 0 & 0 & 0 & 0 & 0 & 1 &0 & 0 &  0 \\
	0 & 0 & 0 & 0 & 0 &  0 &  0 & 0 & 0 & 0 & 0 & 1\\
	[ 0 & 0 & 0 & 0 & 0 & 0 &  0 & 0 &  0 & 0 & 1 & 0. 
	\end{bmatrix}.
	\end{equation}
	Also using equation \eqref{shift}, the shift operator for the complete bipartite graph $K_{(2,3)}$ can be represented by
	\begin{equation}
	S_{K_{(2,3)}} = \begin{bmatrix}
	0 & 0 & 0 & 0 & 0 & 0 & 1 & 0 & 0 & 0 & 0 & 0\\
	0 & 0 & 0 & 0 & 0 & 0 & 0 & 0 & 1 & 0 & 0 & 0\\
	0 & 0 & 0 & 0 & 0 & 0 & 0 & 0 & 0 & 0 & 1 & 0\\
	0 & 0 & 0 & 0 & 0 & 0 & 0 & 1 & 0 & 0 & 0 &  0\\
	0 & 0 & 0 & 0 & 0 & 0 & 0 & 0 & 0 & 1 & 0 & 0\\
	0 & 0 & 0 & 0 & 0 & 0 & 0 & 0 & 0 & 0 & 0 & 1\\
	1& 0 & 0 & 0 & 0 & 0 & 0 & 0 & 0 & 0 & 0 & 0\\
	0 & 0 & 0 & 1 & 0 & 0 & 0 & 0 & 0 & 0 & 0 & 0\\
	0 & 1 & 0 & 0 & 0 & 0 & 0 & 0 & 0 & 0 & 0 & 0\\
	0 & 0 & 0 & 0 & 1 & 0 & 0 & 0 & 0 & 0 & 0 & 0\\
	0 & 0 & 1 & 0 & 0 & 0 & 0 & 0 & 0 & 0 & 0 & 0\\
	0 &0 &0 & 0 & 0 & 1 & 0 & 0 & 0 & 0 & 0 & 0
	\end{bmatrix}.
	\end{equation}
	Therefore, the unitary operator of the DTQW is $U_{K_{(2,3)}} = S_{K_{(2,3)}} \times C_{K_{(2,3)}}$.
	
	The operator $U_{K_{(2,3)}}$ is applied on the sender state $\ket{\psi_0}$, repeatedly in each step of the quantum walk and we calculate $F(\ket{\psi_t}, \ket{\psi_1})$ for $100$ time-steps, which is the fidelity without noise. 		
	The state transfer plots for the complete bipartite graph $K_{(2,3)}$ are shown in subfigures \ref{K_(2,3)_state_transfer_RTN} and \ref{K_(2,3)_state_transfer_OUN} respectively, with and without RTN and OUN noise. From the subfigures, we observe that under non-Markovian RTN, the fidelity plot shows oscillatory behaviour, but under non-Markovian OUN, the fidelity starts decreasing after a few steps.
	
	Now we study the periodicity of the quantum walker at vertex $0$ with and without noise. For the periodicity, we consider $s = r = 0$. It updates the coin operators as 
	\begin{equation}
	C'_{K_{(2,3)}} = \begin{bmatrix}
	0.33 & -0.67 & -0.67 & 0 &  0 & 0 & 0 & 0 & 0 & 0 & 0 & 0\\
	-0.67 & 0.33 & -0.67 &  0 & 0 & 0 & 0 & 0 &  0& 0 & 0 & 0\\
	-0.67 & -0.67 & 0.33 & 0 & 0 & 0 & 0 & 0 & 0 & 0 & 0 & 0\\
	0 & 0 & 0 & -0.33 & 0.67 & 0.67 & 0 & 0 & 0 & 0 & 0 & 0\\
	0 & 0 & 0 & 0.67 & -0.33 & 0.67 & 0 & 0 & 0 & 0 & 0 & 0\\
	0 & 0 & 0 & 0.67 & 0.67 & -0.33 & 0 & 0 & 0 & 0 & 0 &  0\\
	0 & 0 & 0 & 0 & 0 & 0 & 0 & 1 & 0 & 0 & 0 & 0\\
	0 &  0 & 0 & 0 & 0 & 0 & 1 & 0 & 0 & 0 & 0 & 0 \\
	0 & 0 & 0 & 0 & 0 & 0 & 0 & 0 & 0 & 1 & 0 & 0 \\
	0 & 0 & 0 & 0 & 0 & 0 & 0 & 0 & 1 &0 & 0 &  0 \\
	0 & 0 & 0 & 0 & 0 &  0 &  0 & 0 & 0 & 0 & 0 & 1\\
	0 & 0 & 0 & 0 & 0 & 0 &  0 & 0 &  0 & 0 & 1 & 0 
	\end{bmatrix}
	\end{equation} 
	which is different from $C_{K_(2,3)}$ mentioned in equation \eqref{coin_state_transfer_K_{(2,3)}}. 
	
	The periodicity of the walker at vertex $0$ is shown in sub-figures \ref{K_(2,3)_periodicity_RTN} and \ref{K_(2,3)_periodicity_OUN}. The effect of noise can be discerned. The fidelity exhibits an oscillatory pattern under non-Markovian RTN but starts decreasing gradually under non-Markovian OUN noise. This is because of the CP-divisible memory inherent in OUN noise \cite{utagi2020temporal}.  
	\begin{figure}
		\begin{subfigure}{1\textwidth}
			\includegraphics[width=\textwidth]{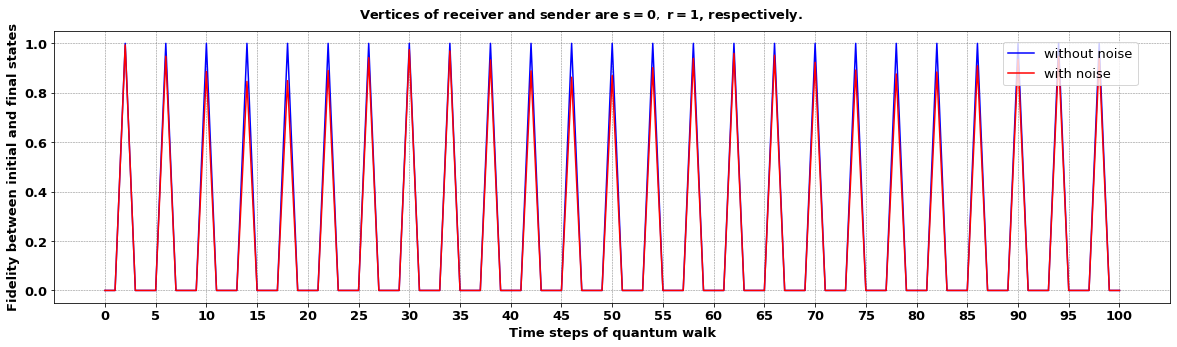}
			\caption{Fidelity with and without RTN noise when both the sender and receiver are located in the same partition.}
			\label{K_(2,3)_state_transfer_RTN} 
		\end{subfigure}
		\begin{subfigure}{1\textwidth}
			\includegraphics[width=\textwidth]{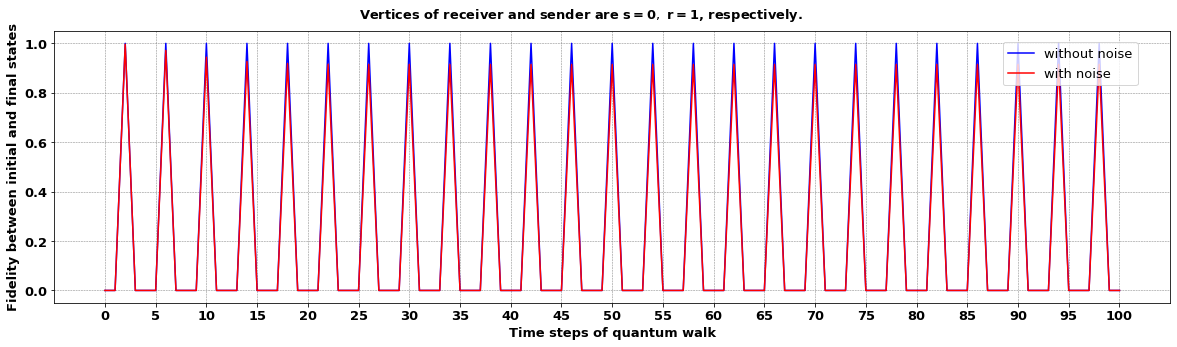}
			\caption{Fidelity with and without OUN noise when both the sender and receiver are located in the same partition.} 
			\label{K_(2,3)_state_transfer_OUN}
		\end{subfigure}
		\begin{subfigure}{1\textwidth}
			\includegraphics[width=\textwidth]{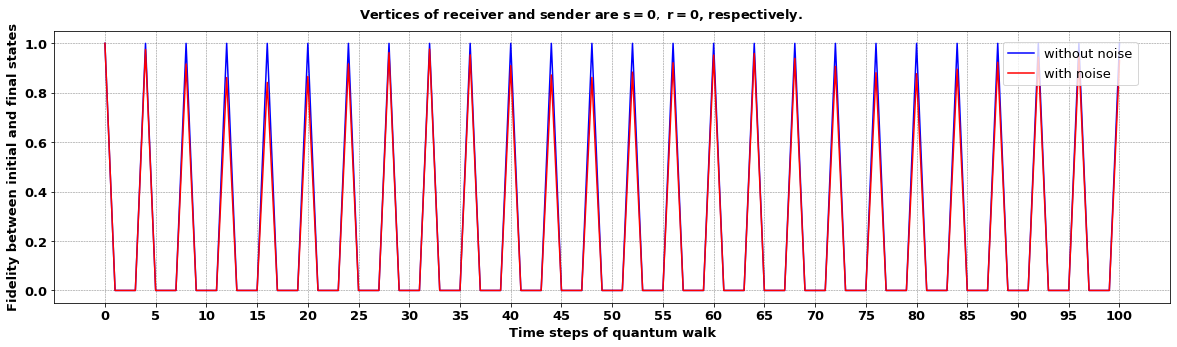}
			\caption{Fidelity with and without RTN noise when both the sender and receiver are located at the central vertex.} 
			\label{K_(2,3)_periodicity_RTN}
		\end{subfigure}
		\begin{subfigure}{1\textwidth}
			\includegraphics[width=\textwidth]{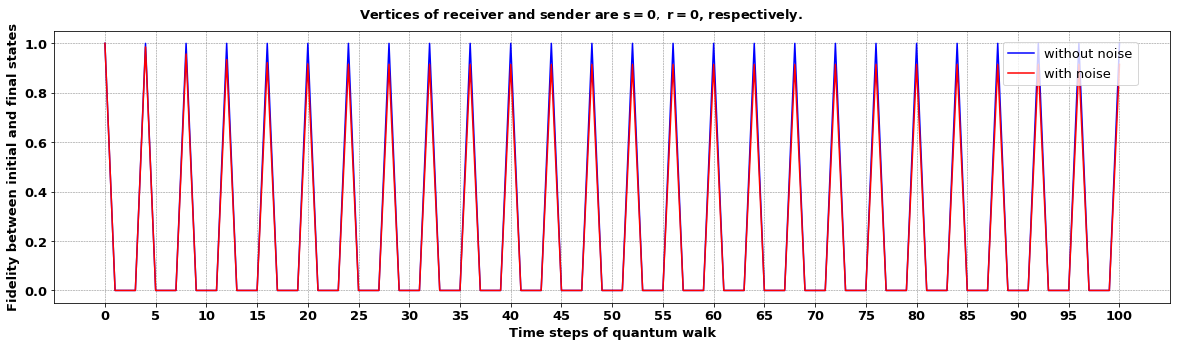}
			\caption{Fidelity with and without OUN noise when both the sender and receiver are located at the central vertex.}
			\label{K_(2,3)_periodicity_OUN}
		\end{subfigure}
		\caption{State-transfer, and Periodicity at vertex $0$ on the complete bipartite graph $K_{(2,3)}$ under non-Markovian RTN and non-Markovian OUN noise. The channel parameters are the same as for the path graph.}
		\label{K_(2,3)_fidelity}
	\end{figure}

	\section{Conclusion}
		
		In this study, we investigated quantum state transfer and periodicity governed by the DTQWs under non-Markovian Random Telegraph Noise (RTN) and modified non-Markovian Ornstein-Uhlenbeck Noise (OUN) across various graph topologies, such as the path graphs, cycle graphs, star graphs, and complete bipartite graphs. We studied the effect of noise on higher-dimensional systems where the Kraus operators were generalized by the Weyl operators. We observed the effect of noise on these graphs by analyzing the fidelity. Different types of behavior of state transfer and periodicity on various graphs were observed. This behaviour was seen to depend on the topology of the network. Our analysis of state transfer under non-Markovian noise would assist in the identification of the graphs and noise regimes where the state transfer fidelity remains high. Since the DTQWs are utilized in search and simulation algorithms, our results indicate that these algorithms may also perform well under noise. Hence, this study sheds new light on various facets of state transfer on different graph topologies under non-Markovian evolution.
        
        Beyond characterizing these dynamics, our study provides practical insights for the design of quantum information protocols. Since DTQWs serve as the basis for search and simulation algorithms, our results suggest that algorithmic performance may remain stable under specific noise regimes and network architectures. Future work could extend this analysis to weighted or time-dependent graphs, explore scalability to larger networks, and investigate the use of revival dynamics as a timing readout for quantum protocols. More broadly, these directions could inform the development of noise-resilient quantum algorithms and architectures leveraging DTQWs under realistic environments.
	
	\section*{Funding}
		This work is supported by the SERB-funded project entitled “Transmission of quantum information using perfect state transfer” (Grant no. CRG/2021/001834).
	
	\section*{Data availability statement}
		All the programs prepared for the numerical calculations are available in GitHub \url{https://github.com/dosupriyo/PST_with_noise}.


\end{document}